\documentclass[aps,pre,epsf,superscriptaddress,amsmath,amssymb,amsfonts,twocolumn,showpacs]{revtex4-1}

\usepackage{graphicx,float}
\usepackage{graphicx}
\usepackage{epsfig}
\usepackage{dcolumn}
\usepackage{bm}
\usepackage{braket}
\usepackage{amsmath}
\usepackage{mathtools}
\usepackage{color}
\usepackage{hyperref}
\newcommand{\abs}[1]{\left| #1 \right|} 

\usepackage{hyperref}
\usepackage{multirow}

\begin{document}
\title{Induced interactions and quench dynamics of bosonic impurities\\ immersed in a Fermi sea}

\author{K. Mukherjee}
\affiliation{Indian Institute of Technology Kharagpur, Kharagpur-721302, West Bengal, India} 
\affiliation{Center for Optical Quantum Technologies, Department of Physics, University of Hamburg, 
Luruper Chaussee 149, 22761 Hamburg Germany}
\author{S. I. Mistakidis}
\affiliation{Center for Optical Quantum Technologies, Department of Physics, University of Hamburg, 
Luruper Chaussee 149, 22761 Hamburg Germany}
\author{S. Majumder}
\affiliation{Indian Institute of Technology Kharagpur, Kharagpur-721302, West Bengal, India} 
\author{P. Schmelcher}
\affiliation{Center for Optical Quantum Technologies, Department of Physics, University of Hamburg, 
Luruper Chaussee 149, 22761 Hamburg Germany}
\affiliation{The Hamburg Center for Ultrafast Imaging, University of Hamburg, Luruper Chaussee 149, 22761 Hamburg, Germany}

\date{\today}


\begin{abstract} 

We unravel the ground state properties and the non-equilibrium quantum dynamics of two bosonic impurities immersed in an 
one-dimensional fermionic environment by applying a quench of the impurity-medium interaction strength. 
In the ground state, the impurities and the Fermi sea are phase-separated for strong impurity-medium repulsions while they experience 
a localization tendency around the trap center for large attractions. 
We demonstrate the presence of attractive induced interactions mediated by the host for 
impurity-medium couplings of either sign and analyze the competition between induced and direct interactions. 
Following a quench to repulsive interactions triggers a breathing motion in both components, with an interaction dependent 
frequency and amplitude for the impurities, and a dynamical phase-separation between the impurities and their surrounding for strong repulsions. 
For attractive post-quench couplings a beating pattern owing its existence to the dominant role of induced interactions takes place 
with both components showing a localization trend around the trap center. 
In both quench scenarios, attractive induced correlations are manifested between non-interacting impurities 
and are found to dominate the direct ones only for quenches to attractive couplings. 

\end{abstract}

\maketitle


\section{Introduction}\label{Introduction} 

Multicomponent quantum gases constitute an ideal playground for investigating a plethora of many-body (MB) 
processes~\cite{Bloch2008,Cazalilla2011} 
including in particular the generation of quasiparticles~\cite{landau1933bewegung,pekar1946autolocalization} such as polarons. 
Quasiparticle formation can be studied owing to the unprecedented experimental tunability of the impurity-medium 
interaction strength, via Feshbach 
resonances~\cite{Ospelkaus2006,Zaccanti2006,chin2010feshbach},  while systems containing few particles 
can be realized especially in one spatial dimension~\cite{serwane2011deterministic,wenz2013few}. 
Depending on the quantum statistics of the host, these quasiparticles are known 
as Bose~\cite{grusdt2015new,rath2013field} and Fermi~\cite{schmidt2018universal,massignan2014polarons} polarons respectively. 
Their existence and a variety of their properties have already been experimentally probed in both 
Bose~\cite{jorgensen2016observation,hu2016bose,yan2020bose} and 
Fermi~\cite{schirotzek2009observation,kohstall2012metastability,koschorreck2012attractive,cetina2016ultrafast,scazza2017repulsive} 
gases, e.g. via employing injection spectroscopy~\cite{cetina2016ultrafast,jorgensen2016observation,hu2016bose}. 
The progress regarding the understanding of the quasiparticles features has also been corroborated by an extensive 
theoretical activity revealing different aspects of their underlying dressing mechanism such as their effective 
mass~\cite{grusdt2017bose,ardila2015impurity}, lifetime~\cite{kohstall2012metastability}, induced 
interactions~\cite{dehkharghani2018coalescence,mistakidis2020many,mistakidis2019repulsive}, and bound 
states termed bipolarons~\cite{camacho2018bipolarons,schmidt2018universal} or trimerons~\cite{nishida2015polaronic,alhyder2020impurity,naidon2018two}. 

Accordingly, the interaction of the impurities with their surrounding leads to deformations of the latter in the vicinity of the former being manifested 
as impurity-medium bound states~\cite{ardila2019strong} for strong attractions as well as sound wave emission~\cite{marchukov2020shape} 
and phase-separation~\cite{mistakidis2020many,mistakidis2019quench} for repulsive interactions. 
These phenomena are a direct imprint of the inevitable entangled nature of these systems~\cite{mistakidis2019quench} 
whose non-equilibrium dynamics is far less appreciated~\cite{skou2020non}. 
The impurity dynamics holds the premise of unveiling even more complex 
processes that will shape our understanding on these settings and may be exploited in future technological applications. 
To date remarkable demonstrations of the impurities' non-equilibrium dynamics include the spontaneous generation 
of nonlinear excitations~\cite{grusdt2017bose,mistakidis2019correlated}, collision induced pattern 
formation~\cite{mistakidis2019dissipative,burovski2014momentum,gamayun2018impact,knap2014quantum,tajima2019collisional}, their 
mediated correlations~\cite{kwasniok2020correlated,mistakidis2020many,mistakidis2020induced} and relaxation 
processes~\cite{mistakidis2020pump,lausch2018prethermalization,boyanovsky2019dynamics}, as well as their 
transport properties in optical lattices~\cite{johnson2011impurity,theel2020entanglement,keiler2020doping,keiler2019interaction}. 
It is also important to emphasize that the above-mentioned investigations have predominantly considered a single impurity 
whilst the effect of larger impurity concentrations leading to enhanced correlation-induced phenomena is until now largely unexplored. 
For these latter settings, the interplay of the quantum statistics between the impurities and the host is of 
importance especially for the induced impurity-impurity correlations. 

In this context, a very promising candidate is a fermionic environment containing two bosonic impurities which can 
interact via direct $s$-wave scattering. 
Indeed, most of the experimental and theoretical endeavors of Fermi polarons have been mainly focused on the 
limiting case of a strongly spin imbalanced Fermi gas ~\cite{Schmidt2012,Vlietinck2013,Mora2009,Trefzger2012,Massignan2013,Pilati2010,Schmidt2011,Sanner2012}, while the situation of bosonic impurities in a Fermi sea is arguably much less studied~\cite{De2014,Fratini2012,cetina2016ultrafast,huber2019medium}. 
In this setting it is very interesting to reveal the presence and nature of induced impurity-impurity interactions which are known to be suppressed for fermionic impurities~\cite{mistakidis2019correlated,dehkharghani2018coalescence,mistakidis2019repulsive}. 
The study of the competition between induced and direct $s$-wave interactions, with the latter being naturally absent for fermionic impurities, is 
an intriguing prospect. 
An additional perspective is the possible emergence of impurity-impurity and impurity-medium bound states for strong attractions. 
Certainly, the identification of the above properties in the dynamical response of the system e.g. subjected to an 
impurity-medium interaction quench~\cite{volosniev2015real,mistakidis2019effective} as well as the characterization of the respective pattern formation especially of the host is desirable. 
In order to address these questions we consider, as a paradigmatic setup, a one-dimensional harmonically trapped Bose-Fermi (BF) mixture consisting of two bosonic impurities immersed in a few-body fermionic environment. 
To track the stationary properties and importantly the quantum dynamics of this impurity setting we resort to the multi-layer multi-configuration time-dependent Hartree method for atomic mixtures (ML-MCTDHX)~\cite{cao2017unified,cao2013multi} being a variational approach that allows us to capture all the relevant correlations of the BF mixture. 

For the ground state we find that the impurities and the fermionic environment 
phase-separate for strong impurity-medium repulsions~\cite{mistakidis2019correlated,Viverit2000,lous2018probing}, while 
they exhibit a localization 
tendency close to the trap center for large attractions. 
Interestingly, attractive induced impurity-impurity interactions~\cite{huber2019medium} mediated by the fermionic host 
are revealed for the case of non-interacting bosons for increasing impurity-medium repulsion and attraction. 
However, for repulsively interacting impurities we unveil that the 
induced interactions dominate the direct $s$-wave ones for increasing impurity-medium attractions. 

A quench from zero to finite repulsive impurity-medium interactions triggers a breathing motion~\cite{boudjemaa2020breathing,kiehn2019spontaneous}, 
in each component, with an interaction dependent frequency and amplitude for the impurities. 
Moreover, a dynamical impurity-bath phase-separation takes place for quenches to strong repulsions. 
Importantly, induced impurity-impurity correlations mediated by the host are identified during the evolution of two 
non-interacting impurities and become more pronounced for quenches to stronger repulsions. 
However, in the case of repulsively interacting impurities a competition of induced and direct interactions is evident 
with the latter dominating the former and enforcing the impurities to reside in a two-body superposition. 

Quenching to attractive impurity-medium interactions gives rise to 
a beating pattern~\cite{mistakidis2020many} on the single-particle level which originates from the participation of two breathing 
frequencies in the dynamics of the impurities due to the dominant presence of their attractive induced interactions. 
The impurities show a spatial localization tendency around the trap center leading to a density accumulation of the Fermi sea at 
their instantaneous location. 
The strength of the attractive induced interactions is larger compared to the reverse quench scenario and it is possible to overcome 
the direct impurities coupling for large post-quench attractions~\cite{mistakidis2020induced,mistakidis2020many}. 
In all cases, we show that the degree of impurity-medium entanglement is appreciable, and exhibits a hierarchy. 
For instance, it is larger for fixed impurity interaction and increasing 
quench amplitude.

This work is structured as follows. 
Section~\ref{theory} introduces the setup under consideration [Sec.~\ref{setup}], the employed many-body variational approach [Sec.~\ref{method}] 
and the main observables [Sec.~\ref{observables}] utilized for the characterization of the ground state and the dynamics of the BF mixture. 
In section~\ref{Ground state} we address the ground state properties of the BF mixture 
with a particular focus on the impurity-impurity induced interactions [Sec.~\ref{corel_ground}]. 
The non-equilibrium dynamics upon considering a quench of the impurity-medium coupling to either repulsive [Sec.~\ref{repulsive_quench}] 
or attractive [Sec.~\ref{attractive quench}] interaction regimes is discussed in Sec.~\ref{quench_dynamics}. 
The emergent entanglement dynamics is presented in Sec.~\ref{entanglemet_dynamics}. 
We summarize our results and give an outlook in Sec.~\ref{conclusion}. 
Appendix~\ref{convergence} elaborates further on the details of our variational method and 
delineates the convergence of the presented results exemplarily.

\section{Theoretical Background}\label{theory} 

\subsection{Setup and Hamiltonian}\label{setup}

We consider a particle-imbalanced ultracold BF mixture containing $N_B=2$ bosonic impurities and $N_F=6$ spin-polarized fermions constituting the environment. 
The mixture is assumed to be mass-balanced i.e. $M_B=M_F\equiv M$ and both species are confined in the same one-dimensional (1D) harmonic trap namely $\omega_B=\omega_F\equiv \omega=0.1$. 
This 1D geometry can be experimentally realized by imposing a strong transverse confinement ($\omega_{\perp}$) compared to the longitudinal ($\omega_{\parallel}$) 
one obeying $\omega=\omega_{\parallel}/\omega_{\perp} \ll 1$~\cite{katsimiga2020observation,serwane2011deterministic,wenz2013few}. 
The individual species of such an approximately mass-balanced BF mixture correspond, for instance, to bosonic and fermionic isotopes of the same element e.g. $^{7}$Li-$^{6}$Li~\cite{Kempen2004, Delehaye2015} or $^{171}$Yb-$^{172}$Yb~\cite{Honda2002}. 
The underlying MB Hamiltonian of the above-described system reads  
\begin{equation}\label{1}
\begin{split}
&H = \sum_{\sigma = F, B}^{}\sum_{i = 1}^{N_\sigma} \bigg [ -\frac{\hbar^2}{2M}\bigg (\frac{\partial}{\partial x^{\sigma}_i} \bigg)^2  
+   \frac{1}{2}M\omega^2(x^\sigma_i)^2 \bigg ] \\ &  + g_{BB}\sum_{ i < j }^{} \delta(x^{B}_i - x^{B}_j) + g_{BF}\sum_{ i = 1 }^{N_F} 
\sum_{j = 1}^{N_B}\delta(x^{F}_i - x^{B}_j).
\end{split}
\end{equation}
Operating in the ultracold regime, $s$-wave scattering constitutes the dominant two-body interaction process and hence interparticle interactions can be modeled by a short-range contact potential~\cite{olshanii1998atomic}. 
Note that for the spin-polarized fermions $s$-wave scattering is forbidden due to the Pauli exclusion principle~\cite{pethick2008bose,pitaevskii2003international} and therefore 
their intraspecies interactions vanish. 
Accordingly, the boson-boson and boson-fermion (alias impurity-medium) 1D effective coupling constants~\cite{olshanii1998atomic} are $g_{BB} = 4 \hbar^2a_{BB}/(M a^2_{\perp}) [1 - C a_{BB}/a^2_{\perp,B}]^{-1}$ and $g_{BF} = 4 \hbar^2a_{BF}/(M a^2_{\perp}) [1 - Ca_{BF}/a^2_{\perp,B}]^{-1}$ respectively. 
Here, $a_{BB}$ ($a_{BF}$) is the three-dimensional boson-boson (boson-fermion) $s$-wave scattering length and $C \approx 1.4603$. 
The parameter $a_\perp = \sqrt{\hbar/M\omega_\perp}$ denotes the transversal confinement length scale, with $\omega_{\perp}$ being the transversal trapping frequency. 
Importantly, the boson-boson and boson-fermion interaction strengths $g_{BB}$ and $g_{BF}$ can be experimentally tuned 
either by means of ${a^s_{BB}}$, ${a^s_{BF}}$ using Feshbach resonances~\cite{kohler2006production,chin2010feshbach} or 
via adjusting ${{\omega_\bot}}$ by employing confinement-induced resonances~\cite{olshanii1998atomic}. 

Below, we rescale the MB Hamiltonian of Eq.~(\ref{1}) in terms of $\hbar \omega_\perp$. 
As a consequence, the length, time and interaction strengths are expressed in units of $\sqrt{\hbar/M\omega_\perp}\equiv a_{\perp}$, $\omega^{-1}_\perp$ and $\sqrt{\hbar^3 \omega_\perp/M}$, respectively. 
It is also worth mentioning that a BF mixture with $N_B \ll N_F$, as the one considered herein, features supressed 
three-body recombination particle losses, since their rate is known 
to be proportional to $N^2_B N_F$~\cite{Helfrich2010}. 

In the following, we characterize the ground-state properties of the highly particle-imbalanced BF mixture particularly focusing on the emergent correlation patterns and unveiling, for instance, phase-separation processes as well as identify impurity-impurity induced interactions for varying boson-boson and impurity-medium interaction strengths, see Sec.~\ref{Ground state}. 
Recall that in the absence of an external confinement the two species are miscible by means that they spatially overlap when $g^2_{BF} < g_{BB}$, otherwise they phase-separate~\cite{mistakidis2018correlation,Viverit2000,Roth2002,lous2018probing,mistakidis2019correlated}. 
In the presence of an external trap and also away from the thermodynamic limit the above-mentioned relation is modified, i.e. $g_{BF}$ should 
become substantially larger than $g_{BB}$ in order to achieve the phase-separation. 
Subsequently, we trigger the non-equilibrium dynamics of the BF mixture by applying a quench of the impurity-medium interaction strength ($g_{BF}$) from zero to either repulsive [Sec.~\ref{repulsive_quench}] or attractive [Sec.~\ref{attractive quench}] couplings. 
Importantly, within these latter post-quench interaction regimes impurity-impurity correlations are finite whilst they vanish for the initial state. 
Thus, the system is driven towards regions of finite impurity-impurity interactions aiming at exploring their dynamical fate, the consequent pattern formation and the associated build-up of correlations.

\subsection{Variational wavefunction ansatz and quantum dynamical approach} \label{method} 

To investigate the ground-state and most importantly the quench dynamics of the particle-imbalanced BF mixture we solve the underlying MB Schr{\"o}dinger 
equation using the variational ML-MCTDHX approach~\cite{cao2017unified,cao2013multi}.   
It is based on expanding the MB wavefunction in terms of a time-dependent and variationally optimized basis. 
This asset enables us to capture both the inter- and the intraspecies correlations of the binary system in a computationally efficient manner compared to methods relying on a time-independent basis set.  

The MB wavefunction, $\Psi_{\rm MB}$, is initially expressed in the form of a truncated Schmidt decomposition of rank $D$~\cite{Horodecki2009}. 
Namely 
\begin{equation} \label{4}
\Psi_{\rm MB}(\vec{x}^B, \vec{x}^F;t) = \sum_{k = 1}^{D} \sqrt{\lambda_k(t)}\Psi^B_k(\vec{x}^B;t)\Psi^F_k(\vec{x}^F;t). 
\end{equation}
The values of the Schmidt coefficients, $\lambda_k(t)$, characterize the degree of entanglement of the binary system. 
In decreasing order they are also known as natural species populations of the $k$-th species function. Evidently, the 
system is entangled~\cite{roncaglia2014,Horodecki2009,mistakidis2018correlation} in the case that more than a single 
coefficients $\lambda_k(t)$ exhibit an non-zero population. 
Then, the many-body state [Eq.~(\ref{4})] is a superposition of the respective species states instead of being a direct product of only two states (non-entangled case). 

As a next step, each of the above-mentioned species functions is expanded in terms of the determinants and permanents of $d_\sigma$ distinct time-dependent fermionic and bosonic single particle functions (SPFs) respectively. 
Therefore, each $\Psi^{\sigma}_k(\vec{x}^{\sigma};t)$ reads  
\begin{equation}\label{5}
\begin{split}
&\Psi_k^{\sigma}(\vec x^{\sigma};t) = \sum_{\substack{l_1,\dots,l_{d_{\sigma}} \\ \sum l_i=N}} C_{k,(l_1,
\dots,l_{d_{\sigma}})}(t)\sum_{i=1}^{N_{\sigma}!} \big ( {\rm sign} (\mathcal{P}_i) \big ) ^{\zeta} \\ & \times \mathcal{P}_i
 \left[ \prod_{j=1}^{l_1} \varphi_1^{\sigma}(x_j;t) \cdots \prod_{j=1}^{l_{d_{\sigma}}} \varphi_{d_{\sigma}}^{\sigma}(x_{K(d_{\sigma})+j};t) \right]. 
  \end{split}
\end{equation} 
In this expression, $C_{k,(l_1,....., l_{d_{\sigma}})}(t)$ denote the time-dependent expansion coefficients of a particular 
determinant for fermions or permanent for bosons and $n_i(t)$ is the occupation number of the SPF, $\varphi_i(x;t)$. 
The index $\zeta = 0, 1 $ for bosons and fermions respectively and $\mathcal{P}$ is the permutation operator exchanging the particle configuration $x_{\nu}^{\sigma}$, $\nu=1,\dots,N_{\sigma}$ within the SPFs. 
Also, $\rm sign(\mathcal{P}_i)$ is the sign of the corresponding permutation and $K(r)\equiv l_1+l_2+\dots+l_{r-1}$,
where $l_{r}$ is the occupation of the $r$-th SPF and $r\in\{1,2,\dots,d_{\sigma}\}$. 
We remark that the bosonic subsystem is termed intraspecies correlated if more than one SPF is occupied otherwise it is fully coherent~\cite{lode2020colloquium}, see also the discussion below. 
On the other hand, the fermionic species exhibit beyond non-trivial Hatree-Fock correlations when more than $N_F$ eigenvalues possess a macroscopic population~\cite{mistakidis2019correlated,erdmann2019phase}. 

The time-evolution of the $(N_B+N_F)$-body wavefunction obeying the MB Hamiltonian of Eq.~(\ref{1}) is determined by calculating the corresponding ML-MCTDHX equations of motion~\cite{cao2013multi}. 
The latter are found by performing e.g. the Dirac-Frenkel variational principle~\cite{Dirac193,Frenkel1934} for the MB ansatz provided by Eqs.~\eqref{4} and \eqref{5}. 
As a result we obtain a set of $D^2$ linear differential equations of motion for the coefficients $\lambda_k(t)$ being coupled to $D[$ ${N_B+d_B-1}\choose{d_B-1}$+${d_F}\choose{N_F}$] non-linear integro-differential equations for the species functions and $d_F+d_B$ integro-differential equations for the SPFs. 
Finally, let us mention in passing that the variational ML-MCTDHX ansatz can be easily reduced to different levels of approximation. 
As a case example, the corresponding mean-field wavefunction ansatz of the BF mixture corresponds to the case of $D = d_B = 1$ and $d_F = N_F$ while the respective mean-field equations of motion are retrieved 
by following a variational principle, see e.g. for details \cite{lode2020colloquium,kohler2019dynamical}.

\subsection{Observables and analysis}\label{observables}

In the following, we briefly introduce the basic observables that will be employed in the remainder of our work in order to characterize 
both the stationary properties and the non-equilibrium dynamics of the BF mixture. 
A particular emphasis is paid on the impurities subsystem. 
To visualize the spatial distribution of the $\sigma=B,F$ species, i.e. the impurities and the medium respectively, on the single-particle 
level we invoke the corresponding one-body reduced density matrix 
\begin{equation}\label{1BD}
\rho^{(1)}_{\sigma}(x, x';t)=\langle \Psi_{\rm MB}(t)| \hat{\Psi}_\sigma^{\dagger}(x) \hat{\Psi}_{\sigma}(x')  | \Psi_{\rm MB}(t) \rangle.
\end{equation}
Here, $\hat{\Psi}_{B}(x)$ [$\hat{\Psi}_{F}(x)$] is the so-called bosonic [fermionic] field operator acting on position $x$ and satisfying the 
standard commutation [anti-commutation] relations~\cite{pethick2008bose,pitaevskii2003international}. 
The diagonal of $\rho^{(1)}_{\sigma}(x, x';t)$ is the well-known one-body density of the $\sigma$-species 
i.e. $\rho^{(1)}_{ \sigma}(x;t) = \rho^{(1)}_ {\sigma}(x, x' = x;t)$~\cite{lode2020colloquium}. 
The latter is accessible in ultracold atom experiments using the single-shot absorption imaging technique~\cite{sakmann2016single,Bloch2008} 
and especially for few atoms can be retrieved by averaging over a sample of single-shots~\cite{mistakidis2018correlation,Klein2005,mistakidis2019dissipative}. 
We remark that the eigenfunctions of the $\sigma$-species one-body reduced density matrix are known as the $\sigma$-species natural orbitals, namely $\phi^{\sigma}_i$. 
In this sense, when more than $N_F$ (one) fermionic (bosonic) natural orbitals are significantly populated the corresponding subsystem is called 
fragmented or intraspecies correlated~\cite{mistakidis2018correlation,lode2020colloquium}. 
Accordingly, the underlying degree of fragmentation can be quantified via measuring $F_F(t) = 1 -\sum_{i = 1}^{N_F}n^F_i(t)$ and $F_B(t) = 1 - n^B_1(t)$ for the fermionic and the bosonic subsystems respectively. 
Here we consider that the population of the total number of orbitals are normalized to unity i.e. 
$\sum_{i=1}^{d_F}n^{F}_i(t) = 1$ and $\sum_{i=1}^{d_B}n^{B}_i(t)=1$. 
Recall that in the MF limit of the BF mixture~\cite{pitaevskii2003international,mistakidis2019correlated,karpiuk2004soliton} where 
$\Psi_{\rm MB}(\vec{x}^F, \vec{x}^B;t) \rightarrow \Psi_{MF}(\vec{x}^{F}, \vec{x}^B ; t) $ the natural populations 
of the fermionic and the bosonic species satisfy the constraints $\sum_{i=1}^{N_F}n^{F}_i(t) = 1$, $n^F_{i > N_F}(t) = 0$, 
and $n^{B}_1(t) = 1$, $n^{B}_{i >1}(t) = 0$. 

The emergence of impurity-medium entanglement can be identified by calculating the Schmidt coefficients, $\lambda_k(t)$, participating in the MB wavefunction ansatz as described by Eq. (\ref{4}). 
Indeed, in the case that more than one coefficients are populated, i.e. $\lambda_{k>1}(t)\neq 0$, then the MB wavefunction is not a single product state and the system is entangled~\cite{Horodecki2009,mistakidis2018correlation}. 
The Schmidt coefficients are essentially the eigenvalues of the species reduced density matrix namely  
$\rho^{N_{\sigma}} (\vec{x}^{\sigma}, \vec{x}'^{\sigma};t)=\int d^{N_{\sigma'}} x^{\sigma'} \Psi^*_{\rm MB}(\vec{x}^{\sigma}, 
\vec{x}^{\sigma'};t) \Psi_{\rm MB}(\vec{x}'^{\sigma},\vec{x}^{\sigma'};t)$, with $\vec{x}^{\sigma}=(x^{\sigma}_1, \cdots,x^{\sigma}_{N_{\sigma-1}})$, and $\sigma\neq \sigma'$. 
Consequently, in order to determine the degree of the impurity-medium entanglement we use the Von-Neumann entropy~\cite{Catani2009, Horodecki2009} given by 
\begin{equation}\label{VN}
S_{VN}(t) = - \sum_{k = 1}^{D} \lambda_k(t) \ln[\lambda_k(t)].
\end{equation}
It becomes apparent that $S_{VN}(t) \geq 0$ only when $\lambda_{k>1}(t)\neq 0$ meaning that entanglement is present. 
For instance, in the mean-field limit where $\lambda_1 (t) = 1$, $\lambda_{k>1}(t)= 0$ and entanglement is absent it holds that $S_{VN}(t)=0$. 

To infer the role of impurity-impurity and fermion-fermion two-body correlation processes in the ground state as well as 
in the dynamics of the BF mixture in a spatially resolved manner we resort to the diagonal of the two-body reduced 
density matrix~\cite{mistakidis2018correlation,lode2020colloquium,sakmann2008reduced}
\begin{equation}\label{2BD}
\begin{split}
 \rho^{(2)}_{\sigma \sigma}(x, x';t) = & \langle \Psi_{\rm MB}(t)| \hat{\Psi}^{\dagger}_{\sigma}(x')
 \hat{\Psi}^{\dagger}_{\sigma}(x) \hat{\Psi}_{\sigma}(x') \\ &
 \times \hat{\Psi}_\sigma(x) | \Psi_{\rm MB}(t) \rangle.
 \end{split}
\end{equation} 
This measure refers to the probability of detecting simultaneously one impurity $\sigma=B$ (fermionic, $\sigma=F$) particle located at $x$ and another one at $x'$. 
In that light, it reveals the occurrence of impurity-impurity (fermion-fermion) two-body correlations and thus provides insights on how 
the two bosons (fermions) behave with respect to one another~\cite{mistakidis2019correlated,erdmann2019phase,mistakidis2020many,mistakidis2020induced}.  

To estimate the strength of the effective interactions between the two bosonic impurities 
we utilize their relative distance~\cite{mistakidis2019correlated,mistakidis2020many,mistakidis2019repulsive} defined as
\begin{equation} \label{7}
\mathcal{D}_{\rm rel}(t) = \frac{\int dx_1 dx_2 \abs{x_1 -x_2} \rho^{(2)}_{BB} (x_1, x_2; t)}{\langle \Psi_{\rm MB}(t) | \hat{N}_B(\hat{N}_B -1) | 
\Psi_{\rm MB}(t) \rangle}.
\end{equation}
Here, $\hat{N}_B$ is the bosonic number operator and $\rho^{(2)}_{BB} (x_1, x_2; t)$ denotes the 
two-body density matrix [Eq.~(\ref{2BD})] of the bosonic impurities subsystem. 
The relative distance can be experimentally accessed using {\it in-situ} spin-resolved single-shot measurements~\cite{bergschneider2018spin}, 
where in particular the actual shape of $\mathcal{D}_{\rm rel}(t)$ can be retrieved by averaging over a sample of the individually obtained images.

\section{Ground state properties of two bosonic impurities in a fermionic environment}\label{Ground state} 

We consider $N_B = 2$ bosonic impurities in a fermionic finite-sized medium composed of $N_F=6$ spin-polarized fermions. 
Recall that an one-dimensional Fermi sea with $N_F>5$ atoms approaches the behavior of a many-body fermionic environment, see for instance 
Ref.~\cite{wenz2013few} for a corresponding experimental verification. 
In our setting we have checked that our results, to be presented below, regarding both the ground state and the dynamics remain qualitatively the same also for $N_F=8$ (not shown here for brevity). 
The system is mass-balanced and both species are trapped in the same harmonic oscillator of frequency $\omega = 0.1$, unless it is stated otherwise. 
Below, we examine the ground state characteristics of the composite system with a particular focus on the impurities properties for attractive and repulsive impurity-medium interactions. 
In order to discriminate between direct and effective impurity-impurity interaction effects we analyze both the cases of non-interacting and interacting impurities. 
The impact of the impurities mass on their induced interactions mediated by the fermionic environment is also discussed. 
Another objective of our analysis is to unveil the spatial distributions of each species, discuss possibly emerging phases of the BF mixture as well as their associated correlation properties for varying impurity-medium interactions. 
To obtain the ground state of the BF mixture governed by Eq.~\eqref{1} we employ either the imaginary-time propagation or the improved relaxation 
method within ML-MCTDHX~\cite{cao2017unified}.

\subsection{Single-particle density distribution}\label{ground_state_density} 

\begin{figure}[H]
\centering
\includegraphics[width=0.45\textwidth]{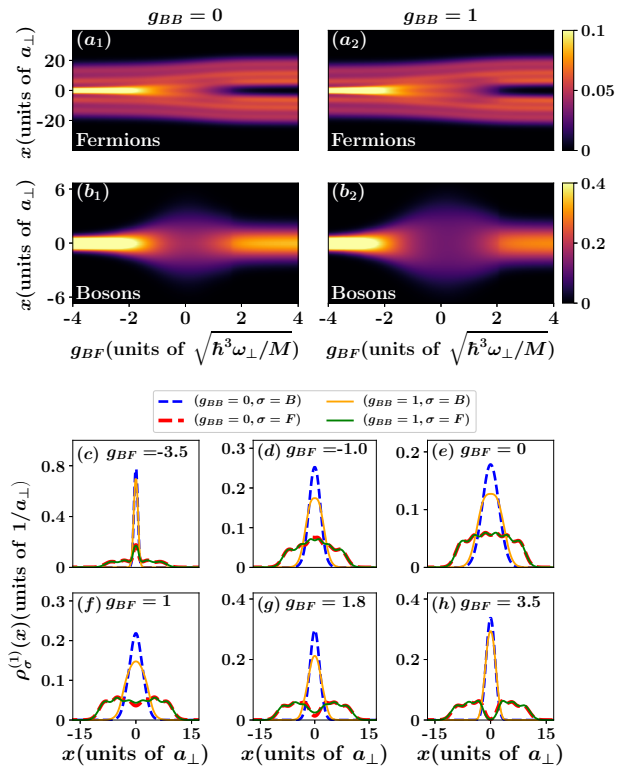}
\caption{Ground state single-particle density of ($a_1$), ($a_2$) the six fermions and ($b_1$),($b_2$) the two bosons for varying impurity-medium interaction strength $g_{BF}$ when [($a_1$), ($b_1$)] $g_{BB}=0$ and [($a_2$), ($b_2$)] $g_{BB}=1$. 
($c$)-($h$) Density profiles $\rho^{(1)}_{B}(x)$ and $\rho^{(1)}_{F}(x)$ at specific $g_{BF}$ values and for both non-interacting and repulsively interacting impurities (see legend). 
The BF mixture consists of $N_B = 2$ bosonic impurities and $N_F = 6$ fermions both trapped in a harmonic oscillator with $\omega = 0.1$.}  
\label{spd_g}
\end{figure}

Let us first inspect the spatial configuration of the ground state of the bosonic impurities and the fermionic sea for varying impurity-medium interaction strength $g_{BF}$. 
For this reason, we employ the corresponding single-particle densities $\rho^{(1)}_{F}(x)$ and $\rho^{(1)}_{B}(x)$ with respect to $g_{BF}$ [Fig.~\ref{spd_g}] 
for both the cases of two non-interacting $g_{BB}=0$ [Figs.~\ref{spd_g}($a_1$), ($b_1$)] and two repulsively interacting 
with $g_{BB}=1$ [Figs. \ref{spd_g} ($a_2$), ($b_2$)] bosonic impurities. 
Overall, we observe that the behavior of both the impurities and the medium depends strongly on the value of $g_{BF}$. 
Also $\rho^{(1)}_{F}(x)$ exhibits six shallow local density maxima [Figs.~\ref{spd_g}($c$)-($h$)] almost irrespectively of $g_{BF}$, which 
indicates the presence of six fermions~\cite{kwasniok2020correlated}, see 
also the remark in \footnote{Note that for increasing attraction, i.e. $g_{BF}<-2.5$, a major portion of $\rho^{(1)}_{F}(x)$ resides around 
$x=0$ and its two central local maxima come very close to each other and eventually merge for very strong attractions.}. 
Interestingly, the shape of $\rho^{(1)}_{F}(x)$ for fixed $g_{BF}$ remains almost unchanged between the $g_{BB}=0$ and the $g_{BB}=1$ cases, 
see Figs.~\ref{spd_g}($a_1$) and ($a_2$) as well as Figs.~\ref{spd_g}($c$)-($h$). 
On the other hand, $\rho^{(1)}_{B}(x)$ at a certain value of $g_{BF}$ is affected by the direct impurity-impurity interactions since 
for $g_{BB}=1$ it becomes slightly broader than for $g_{BB}=0$ especially when $-1.5<g_{BF}<1.5$, see Figs.~\ref{spd_g}($d$)-($g$). 
As we shall argue below, this difference is caused by the presence of attractive induced impurity-impurity interactions mediated by the Fermi sea and become more 
pronounced when $g_{BB}=0$. 
A similar effect has also been discussed in the context of two bosonic impurities in a BEC bath~\cite{mistakidis2020many,mistakidis2020pump}. 
The impact of $g_{BB}$ on the behavior of the impurities and the interplay between direct and  attractive impurity-impurity induced interactions 
will be discussed in detail in Sec.~\ref{corel_ground}.  

In particular, for $-1.5<g_{BF}<1.5$ the Fermi sea exhibits a broad spatial distribution as identified by $\rho^{(1)}_{F}(x)$ enclosing the 
bosonic impurities whose $\rho^{(1)}_{B}(x)$ is located around the trap center [Figs.~\ref{spd_g}($d$)-($f$)]. 
Therefore, for these impurity-medium interactions the bosons and the Fermi sea show a miscible behavior~\cite{mistakidis2019correlated} independently of $g_{BB}$, 
see Figs. \ref{spd_g} ($a_1$), ($a_2$). Turning to attractive $g_{BF}<-1.5$ we observe a drastically altered behavior of both the impurities 
and the fermionic ensemble. 
More specifically, the impurities are strongly localized in the vicinity of the trap center since $\rho^{(1)}_{B}(x)$ shows a peaked structure 
at $-1<x<1$ having a sech-like shape [Fig.~\ref{spd_g}($b_1$), ($b_2$)]. 
This spatial localization tendency of the impurities signals the formation of an attractively bound pair~\cite{mistakidis2020many,mistakidis2020pump} 
as a consequence of the presence of strong attractive induced impurity-impurity interactions \cite{mistakidis2020many,mistakidis2020induced}, 
see for more details Sec.~\ref{corel_ground}. 
Simultaneously $\rho^{(1)}_{F}(x)$ majorly accumulates at the same location by developing a density hump, see for instance 
Figs.~\ref{spd_g}($a_1$), ($a_2$) and ($c$), while its background is significantly reduced when compared to smaller $g_{BF}$ values. 
The latter behavior of $\rho^{(1)}_{F}(x)$ indicates that the fermions of the medium lie very close to the impurities suggesting 
that for even larger attractions a bound state between an impurity and an atom of the Fermi sea may be formed~\cite{mistakidis2019dissipative}, a phenomenon which is not further examined herein. 
Of course, the fermions of the medium can not reside exactly at the same location due to the Pauli exclusion 
principle~\cite{pethick2008bose,pitaevskii2003international}, see also the discussion in Sec.~\ref{corel_ground}. 

On the other hand, for strong repulsive impurity-bath interactions such that $g_{BF}>1.5$ the spatial configuration of the system and especially of the Fermi sea is significantly changed compared to smaller values of $g_{BF}$. 
Indeed, a local density dip builds upon $\rho^{(1)}_{F}(x)$ around $x\approx 0$ [Fig.~\ref{spd_g}($g$)] which becomes more pronounced for increasing $g_{BF}$ and for $g_{BF}>2$ $\rho^{(1)}_{F}(x)$ is segregated into two fragments residing in the left and right side with respect to $x=0$ [Figs. \ref{spd_g} ($a_1$), ($h$)]. 
Note that each of the fragments has three local density maxima indicating that predominantly three fermions populate each of them and also reflects the fact that the first six lowest-lying single-particle eigenstates of the harmonic trap majorly contribute to the fermionic MB wavefunction. 
The impurities density $\rho^{(1)}_{B}(x)$ lies in between the two fragments of $\rho^{(1)}_{F}(x)$ and therefore an impurity-medium phase-separation process takes place~\cite{mistakidis2018correlation}, see e.g. Figs. \ref{spd_g} ($a_1$), ($a_2$) and ($h$). 
This procedure is identified by the small spatial overlap among the components~\cite{mistakidis2018correlation} which becomes suppressed for increasing $g_{BF}$. 
We remark that the phase-separation region is shifted to larger $g_{BF}$ values when $g_{BB}$ is finite, compare in particular Figs. \ref{spd_g} ($a_1$) and ($a_2$). 
Indeed, phase-separation occurs when the interspecies interaction energy overcomes the intraspecies one and thus a larger $g_{BF}$ is required for 
increasing $g_{BB}$~\cite{mistakidis2018correlation} in order to accomplish this process. 
It is also worth mentioning at this point that a system of two fermionic impurities immersed in a bosonic bath exhibits a similar phase-separation behavior at repulsive $g_{BF}$ but in this case the impurities reside at the edges of 
the bosonic medium~\cite{mistakidis2019correlated}. 

The above-described phase-separation process as well as the localization tendency of the components taking place at large repulsive and attractive impurity-medium interactions respectively can be intuitively understood in terms of an effective potential approach~\cite{mistakidis2020many,mistakidis2019quench,kiehn2019spontaneous}. 
For this picture one can consider an effective potential for the impurities [Fermi sea] constructed by superimposing the single-particle density of the Fermi sea [impurities] to the external harmonic trap, namely $V_{\rm eff}^{B}(x) = \frac{1}{2}m \omega^2 x^2 + g_{BF}\rho^{(1)}_{F}(x)$ [$V_{\rm eff}^{F}(x) = \frac{1}{2}m \omega^2 x^2 + g_{BF}\rho^{(1)}_{B}(x)$]. 
Referring to the impurities subsystem at strong repulsive $g_{BF}$ their effective potential $V_{\rm eff}^{B}(x)$ corresponds to a deformed harmonic trap due to $\rho^{(1)}_{F}(x)$, see e.g. Fig. \ref{spd_g} ($h$). 
In this sense the impurities reside around the trap center possessing a Gaussian-like spatial distribution [Fig.~\ref{spd_g}($h$)]. 
On the other hand, for $g_{BF}>0$ the corresponding $V_{\rm eff}^{F}(x)$ has a double-well like structure where the role of the potential barrier at $x=0$ is played by $\rho^{(1)}_{B}(x)$. 
In turn, this $V_{\rm eff}^{F}(x)$ enforces the splitting of $\rho^{(1)}_{F}(x)$ into two fragments, see e.g. Fig.~\ref{spd_g}($h$). 
Note also here that for $g_{BB}=1$ the maximum of $\rho^{(1)}_{B}(x)$ is smaller compared to the $g_{BB}=0$ case [Fig.~\ref{spd_g}($g$)]. 
This gives rise to a shallower double-well effective potential for fixed $g_{BF}$ and thus the barrier height that allows for phase-separation is achieved for larger values of $g_{BF}$ when $g_{BB}$ is finite. 
A similar argumentation can also be applied for attractive $g_{BF}$ where, for instance, the aforementioned localization tendency of $\rho^{(1)}_{B}(x)$ is essentially determined by the hump structure building upon $\rho^{(1)}_{F}(x)$ [Fig. \ref{spd_g} ($c$)] and vice versa due to back-action. 
For more details on the range of applicability of this effective potential picture we refer the interested 
reader to Refs.~\cite{mistakidis2020many,mistakidis2019dissipative,mistakidis2019quench,kiehn2019spontaneous}.

\subsection{Impurity-impurity induced interactions}\label{corel_ground}

\begin{figure}[ht]
\centering
\includegraphics[width=0.45\textwidth]{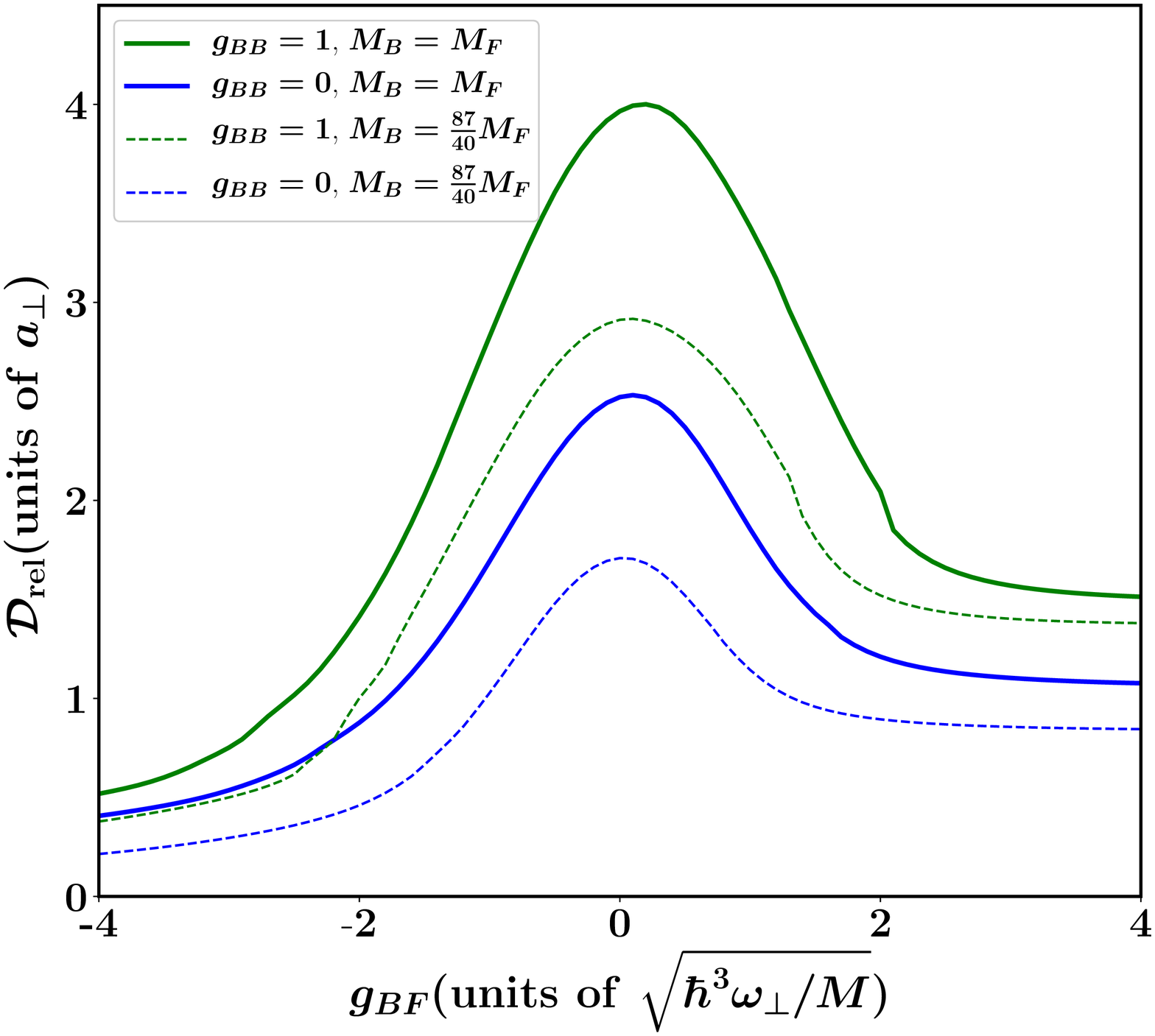}
\caption{Relative distance $\mathcal{D_{\rm rel}}$ between the two bosonic impurities in the ground state of the BF mixture with 
respect to the impurity-medium interaction strength $g_{BF}$. The relative distance is presented for the cases of two 
non-interacting ($g_{BB} = 0$), and  interacting ($g_{BB} = 1$) impurities in a mass-balanced as well as a mass-imbalanced system (see legend). 
The medium consists of $N_F = 6$ fermions, while the BF mixture is confined in a harmonic trap with frequency $\omega = 0.1$.}  
\label{Relative_dis}
\end{figure}

\begin{figure}[ht]
\centering
\includegraphics[width=0.47\textwidth]{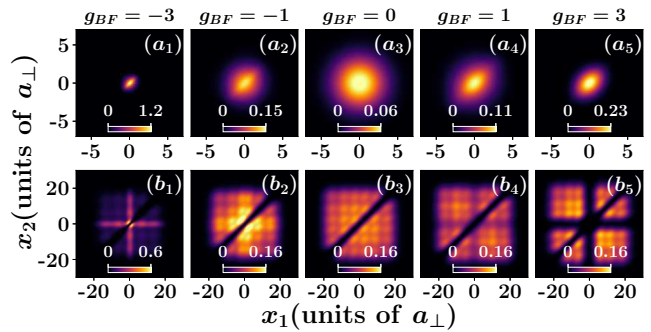}
\caption{Reduced two-body ($a_1$)-($a_5$) boson-boson $\rho^{(2)}_{BB}(x_1, x_2)$, and ($b_1$)-($b_5$) fermion-fermion $\rho^{(2)}_{FF}(x_1, x_2)$ 
density in the ground state of the BF mixture for selective impurity-medium interaction strengths (see legend).  
The system contains $N_B =2$ non-interacting ($g_{BB}=0$) bosonic impurities and $N_F = 6$ fermions. 
It is further confined in a harmonic trap of $\omega = 0.1$. }  
\label{2bd_g0}
\end{figure}

\begin{figure}[ht]
\centering
\includegraphics[width=0.47\textwidth]{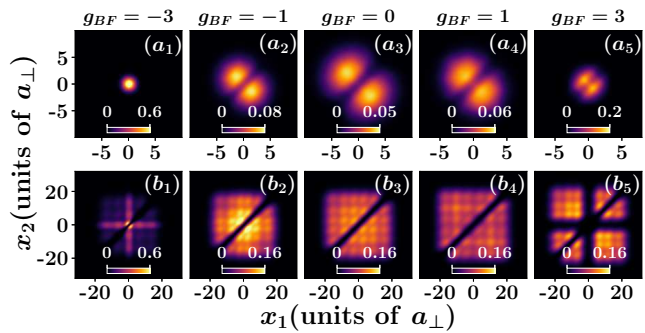}
\caption{Reduced two-body ($a_1$)-($a_5$) impurity-impurity $\rho^{(2)}_{BB}(x_1, x_2)$, and ($b_1$)-($b_5$) fermion-fermion $\rho^{(2)}_{FF}(x_1,x_2)$ 
density in the ground state of the BF mixture for different values of the boson-medium coupling constant $g_{BF}$ (see legend). 
The system consists of $N_B =2$ interacting with $g_{BB}=1$ bosonic impurities and $N_F = 6$ fermions while it is trapped in a harmonic oscillator with frequency $\omega = 0.1$.}  
\label{2bd_g1}
\end{figure}

The impurities being immersed in the Fermi sea are dressed by its excitations forming quasiparticles, herein Fermi 
polarons~\cite{schmidt2018universal,massignan2014polarons,mistakidis2019repulsive}. 
An intriguing property of the generated quasiparticles is the emergence of attractive induced interactions among them mediated by their host~\cite{dehkharghani2018coalescence,mistakidis2020many,mistakidis2019repulsive} and that they can possibly form a bound pair for strong impurity-medium attractions~\cite{schmidt2018universal,massignan2014polarons}. 
To identify such quasiparticle related mechanisms in the ground state of the BF mixture we subsequently inspect the relative 
distance $\mathcal{D}_{\rm rel}$ [Eq.~(\ref{7})] and the spatially resolved two-body reduced density matrix $\rho^{(2)}_{BB}(x_{1}, x_2)$ of 
the bosonic impurities~\cite{mistakidis2020induced} for different impurity-medium interaction strengths, see 
Figs.~\ref{Relative_dis}, \ref{2bd_g0} and \ref{2bd_g1}. 
Recall that $\rho^{(2)}_{BB}(x_1,x_2)$ measures the probability of finding a boson at position $x_1$ while the second one is located at $x_2$. 
Importantly, the combination of the behavior of $\mathcal{D}_{\rm rel}$ and $\rho^{(2)}_{BB}(x_{1}, x_2)$ enables us to infer the presence 
and strength of the attractive induced interactions as well as the spatial configuration of the impurities~\cite{mistakidis2020many}.   
Below, we discuss the cases of both non-interacting ($g_{BB}=0$) and repulsively interacting ($g_{BB}=1.0$) impurities as well as the 
effect of a mass-imbalance between the impurities and their bath. 

The corresponding relative distance $\mathcal{D}_{\rm rel}$ between the impurities and their two-body density $\rho^{(2)}_{BB}(x_{1}, x_2)$ 
for a mass-balanced BF system containing non-interacting impurities are presented in Fig.~\ref{Relative_dis} and 
Figs.~\ref{2bd_g0}($a_1$)-($a_5$) respectively as a function of the impurity-medium interaction strength $g_{BF}$ ranging 
from attractive to repulsive values. 
It becomes apparent that $\mathcal{D}_{\rm rel}$ gradually decreases, when compared to its value for $g_{BF}=0$, as $\abs{g_{BF}}$ is 
increased towards the attractive or the repulsive interaction regime. This overall decreasing behavior of $\mathcal{D}_{\rm rel}$ for 
larger $\abs{g_{BF}}$ indicates that for finite attractive or repulsive impurity-medium interactions the impurities move close to each other as compared to the $g_{BF}=0$ scenario. 
The latter tendency suggests the emergence of attractive impurity-impurity induced interactions~\cite{huber2019medium,mistakidis2020many}. 
Interestingly, $\mathcal{D}_{\rm rel}$ tends to approach a constant value which is different for strong repulsions ($g_{BF}>3$) and 
attractions ($g_{BF}<-3$). 
Indeed, the saturation value of $\mathcal{D}_{\rm rel}$ for strong repulsions is somewhat larger when compared to the corresponding 
value for strong attractions. 
This means that for attractive $g_{BF}$ the impurities are substantially closer with respect to one another than in the repulsive case. 
Concluding, $\mathcal{D}_{\rm rel}$ signals the presence of induced impurity-impurity interactions, which are manifested to be attractive 
in general, irrespectively of the sign of the impurity-medium coupling~\cite{huber2019medium,mistakidis2020many}.

To confirm the existence of attractive impurity-impurity induced interactions when $g_{BB}=0$ we next rely on the impurities two 
particle density $\rho^{(2)}_{BB}(x_1, x_2)$, see Figs.~\ref{2bd_g0} ($a_1$)-($a_5$). 
This quantity allows us to explicitly identify the spatial distribution of impurities. 
As it can be seen, irrespectively of the value of $g_{BF}$ the two non-interacting bosons prefer to reside together close to the 
trap center since $\rho^{(2)}_{BB}(x_1, x_2)$ shows a maximum value in the vicinity of $x_1=0$, $x_2=0$, see Figs.~\ref{2bd_g0}($a_1$)-($a_5$). 
In particular, for $g_{BF} = 0$ [Fig.~\ref{2bd_g0}($a_3$)] $\rho^{(2)}_{BB}(x_1, x_2)$ has a circularly symmetric shape in the ($x_1,x_2$)-plane while showing a peak around $x_1=x_2=0$. 
This can be understood by the fact that in the absence of any correlation with the majority species there is no induced interaction among the bosons. 
Hence, the probability of finding the two bosons together at $x_1=x_2$ or one at $x_1$ and the other at $x_2=-x_1$ is the same and becomes maximal at the trap minimum i.e. at $x_1=x_2=0$. 
However, for a finite $g_{BF}$ the shape of $\rho^{(2)}_{BB}(x_1, x_2)$ is significantly altered when compared to the $g_{BF}=0$ case 
since predominantly the diagonal $\rho^{(2)}_{BB}(x_1,x_2=x_1)$ is populated. 
In fact, $\rho^{(2)}_{BB}(x_1, x_2)$ becomes more elongated along the diagonal ($x_1 = x_2$) with increasing $\abs{g_{BF}}$, while it 
shrinks across its anti-diagonal ($x_2 = -x_1$), see Figs.~\ref{2bd_g0}($a_4$)-($a_5$) and Figs.~\ref{2bd_g0}($a_1$)-($a_2$). 
This means that the probability of detecting the two bosons at two different positions is substantially smaller than that of being 
close together. 
Therefore, an effective attractive interaction between the impurities is established and occurs for both attractive and repulsive 
impurity-medium interactions. 
Importantly, within the attractive $g_{BF}$ regime the shrinking of the anti-diagonal of $\rho^{(2)}_{BB}(x_1, x_2)$ is much more 
pronounced than on the repulsive $g_{BF}$ side, compare in particular Figs.~\ref{2bd_g0}($a_1$)-($a_2$) with Figs.~\ref{2bd_g0}($a_4$)-($a_5$). 
The latter observation supports our previous statement in terms of $\mathcal{D}_{\rm rel}$ of a much stronger effective attraction 
between the impurities for negative than positive $g_{BF}$, a result that also holds for Bose polarons as it has been 
demonstrated in Ref.~\cite{mistakidis2020many}. 
Moreover, the pronounced elongation along the diagonal accompanied by the strong suppression of the anti-diagonal of $\rho^{(2)}_{BB}(x_1, x_2)$ e.g. for $g_{BF}=-3$ is indicative of a bound state being formed between the impurities known as a bipolaron state~\cite{mistakidis2020many,camacho2018bipolarons,schmidt2018universal}.     

Next, we turn our attention to repulsively interacting bosonic impurities with $g_{BB}=1$ aiming to investigate the competition between attractive induced interactions and direct $s$-wave ones. 
To this end, we measure the impurities relative distance [Fig.~\ref{Relative_dis}] and their two-particle density [Fig.~\ref{2bd_g1}($a_1$)-($a_5$)] for distinct values of $g_{BF}$. 
As expected, in the absence of direct impurity-impurity interactions i.e. $g_{BB}=0$ the impurities distance $\mathcal{D}_{\rm rel}$ is in general smaller than the corresponding for two repulsively interacting ones with $g_{BB} = 1$. 
This difference becomes maximal for zero impurity-medium interactions, namely $g_{BF}=0$. 
Indeed, $\mathcal{D}_{\rm rel}$ decreases for a larger positive or negative $g_{BF}$ tending to approach a constant value which is smaller for attractive $g_{BF}$ interactions. 
Consequently, also the difference $\mathcal{D}_{\rm rel}(g_{BB} = 1)-\mathcal{D}_{\rm rel}(g_{BB}=0)$ gradually decreases and becomes constant for increasing $\abs{g_{BF}}$. 
A direct comparison between $\mathcal{D}_{\rm rel}(g_{BB}=1)$ and $\mathcal{D}_{\rm rel}(g_{BB}=0)$ reveals that the 
distance saturates at relatively larger (smaller) positive (negative) $g_{BF}$ values when $g_{BB}=1$. 
For instance, the decreasing rate of $\mathcal{D}_{\rm rel}$ in the attractive $g_{BF}$ regime is much larger in the $g_{BB}=1$ scenario before showcasing a saturation tendency around $g_{BF}=-4$ towards $\mathcal{D}_{\rm rel} \approx 0.5$. 
The above-described overall behavior of $\mathcal{D}_{\rm rel}$ for varying $g_{BF}$ suggests the occurence of attractive induced interactions for a finite $g_{BF}$ despite the existence of direct $s$-wave ones. 
Nevertheless, as we shall explicate in the following the direct interactions dominate the induced ones at least for repulsive impurity-medium 
couplings, a result which reveals that the strength of induced interactions is larger in the negative $g_{BF}$ regime.   

Indeed by inspecting the impurities two-body density $\rho^{(2)}_{BB}(x_1,x_2)$, illustrated in Figs.~\ref{2bd_g1}($a_1$)-($a_5$), the following conclusions can be immediately drawn. 
The circularly symmetric pattern of $\rho^{(2)}_{BB}(x_1,x_2)$ occurring for $g_{BB}=0$ when $g_{BF}=0$ [Fig.~\ref{2bd_g0}($a_3$)] is completely modified for repulsively interacting impurities [Fig.~\ref{2bd_g1}($a_3$)]. 
This modification favors a pattern whose diagonal is depleted, giving rise to a correlation hole~\cite{erdmann2019phase,kwasniok2020correlated}, whilst the anti-diagonal develops two symmetric lobes with respect to $x_1=x_2$ and it is predominantly populated. 
This is an explicit imprint of the direct $s$-wave interaction among the impurities and means that the probability of finding two bosons exactly at the same position is vanishingly 
small in contrast to the situation where each boson resides on a separate side in terms of the trap center. 
Switching on $\abs{g_{BF}}$ introduces deformations in the shape of $\rho^{(2)}_{BB}(x_1, x_2)$ and in particular in the position of its anti-diagonal lobes which suggests that the induced interactions set in. 
Referring to repulsive impurity-medium interactions [Figs.~\ref{2bd_g1}($a_4$-($a_5$)] it is apparent that for increasing $g_{BF}$ the anti-diagonal lobes of $\rho^{(2)}_{BB}(x_1, x_2)$ approach the diagonal. 
Therefore, the two bosons get closer due to the presence of their attractive induced interactions mediated by the fermionic environment. 
Notice, however, that the two lobe structure of $\rho^{(2)}_{BB}(x_1, x_2)$ is maintained also at $g_{BF}=3$ [Fig.~\ref{2bd_g1}($a_5$)], indicating that the $s$-wave interactions dominate the induced ones. Turning to the attractive $g_{BF}$ regime we observe that for weak $g_{BF}$ values the anti-diagonal of $\rho^{(2)}_{BB}(x_1, x_2)$ shrinks and thus the bosons come closer when compared to the $g_{BF}=0$ case due to the existence of attractive induced interactions. 
Importantly, this behavior of $\rho^{(2)}_{BB}(x_1, x_2)$ is drastically changed for large attractive impurity-medium interactions. 
More precisely, the two-lobed anti-diagonal structure related to the dominant repulsive contact interaction is changed into a circularly symmetric pattern, see e.g. $\rho^{(2)}_{BB}(x_1,x_2)$ at $g_{BF}=-3$ depicted in Fig.~\ref{2bd_g1}($a_1$). 
Recall that the appearance of such a circularly symmetric structure in $\rho^{(2)}_{BB}(x_1, x_2)$ occurs in the case of zero effective interactions between the two bosons when $g_{BB}=0$ and $g_{BF}=0$ [Fig.~\ref{2bd_g0}($a_3$)]. 
This observation suggests that the attractive induced interactions nullify the direct repulsive contact ones for large attractive impurity-medium couplings, a phenomenon that is absent in the repulsive $g_{BF}$ regime. 
Summarizing, attractive impurity-medium couplings lead to stronger induced interactions than repulsive ones. 

Subsequently, we study the impact of the impurities mass on the strength of the induced interactions by  invoking as an appropriate measure the impurities relative distance [Fig. \ref{Relative_dis}]. 
For this investigation we consider a harmonically trapped mass-imbalanced BF mixture consisting of a $^{40}$K fermionic environment and two $^{87}$Rb bosonic impurities~\cite{Fratini2012}. 
Evidently, the overall phenomenology of $\mathcal{D}_{\rm rel}$ for varying $g_{BF}$ is similar to the mass-balanced scenario for both the $g_{BB}=0$ and the $g_{BB}=1$ cases. 
Moreover, $\mathcal{D}_{\rm rel}$ is always reduced when compared to the mass-balanced system, thus suggesting that heavier impurities prefer to stay closer to each other than lighter ones~\cite{kwasniok2020correlated,mistakidis2020many}. Accordingly, we can deduce that an 
increasing impurity mass allows for stronger attractive induced interactions.

\subsection{Two-body correlations of the fermionic medium}\label{corel_ground_bath} 

Having explicated the existence of attractive induced impurity-impurity correlations we then analyze 
the two-particle distributions of the fermionic environment for different impurity-medium interactions. 
Our main objective here is to expose the back-action of the impurities onto their host~\cite{mukherjee2020pulse,mistakidis2019dissipative}. 
Regarding the system with two non-interacting impurities ($g_{BB}=0$) $\rho^{(2)}_{FF}(x_1,x_2)$ is presented in Figs. \ref{2bd_g0} ($b_1$)-($b_5$) for specific values of $g_{BF}$. 
A depleted diagonal is observed irrespectively of $g_{BF}$ due to the Pauli exclusion principle, namely two fermions cannot occupy the same spatial region. 
At $g_{BF}=0$ two fermions can be found at any two distinct positions within the interval $[-10, 10]$, see Fig.~\ref{2bd_g0}($b_3$), possessing a slightly larger probability to reside close to the trap center either on the same side or at different ones with respect to $x=0$. 
Interestingly, even the presence of a very small number of bosonic impurities is able to significantly alter the properties of the Fermi sea if $g_{BF}\neq 0$. 
For repulsive $g_{BF}$, the fermions exhibit a tendency to stay away from the trap center, 
e.g. $\rho^{(2)}_{FF}(x_1=5,x_2=-5)\approx 0.14$ at $g_{BF} = 1$ in Fig. \ref{2bd_g0} ($b_4$). 
This behavior is manifested by the appearance of relatively low density stripes along the lines $x_1=0$ and $x_2=0$ e.g. for $g_{BF}=1$ [Fig.~\ref{2bd_g0}($a_4$)] which are transformed into completely depleted density regions e.g. for $g_{BF} = 3$ [Fig. \ref{2bd_g0}($a_5$)]. 
Turning to the attractive $g_{BF}$ regime [Figs.~\ref{2bd_g0}($b_1$)-($b_2$)], the distribution of $\rho^{(2)}_{FF}(x_1,x_2)$ is changed significantly. 
Indeed, the probability of finding two fermions at different positions in the vicinity of the trap center is the dominant contribution to $\rho^{(2)}_{FF}(x_1,x_2)$ especially for larger attractions. 
For instance, at $g_{BF} = -1$, the two particle density shown in Fig.~\ref{2bd_g0}($b_2$) is higher close to the trap 
center [e.g. $\rho^{(2)}_{FF}(x_1=1,x_2=-1)\approx 0.16$] compared to the edges [e.g. $\rho^{(2)}_{FF}(x_1=5,x_2=-14)\approx 0.09$]. 
Also, for $g_{BF} = -3$, the spatial region apart from the one close to the trap center is almost completely depleted [see Fig.~\ref{2bd_g0}($b_1$)], as identified by the relevant cross-like pattern building upon $\rho^{(2)}_ {FF}(x_1,x_2)$. 

Comparing now $\rho^{(2)}_{FF}(x_1, x_2)$ between the cases of $g_{BB}=1$ [Figs.~\ref{2bd_g1}($b_1$)-($b_5$)] and $g_{BB}=0$ [Figs.~\ref{2bd_g0}($b_1$)-($b_5$)] we can easily deduce that their shapes at specific $g_{BF}$ values are to a great extent similar. 
A slight difference occurs for moderate repulsive interactions e.g. $g_{BF}=1$ where the two-body density stripes imprinted along the lines $x_1=0$ and $x_2=0$ for $g_{BB}=0$ [Fig.~\ref{2bd_g0}($b_4$)] 
are not noticeable for $g_{BB}=1$ [Fig.~\ref{2bd_g1}($b_4$)]. 
Also, for attractive impurity-medium interactions $g_{BF}<0$ the regions away from the trap center are relatively 
stronger populated for $g_{BB}=1$, compare Figs. \ref{2bd_g1} ($b_1$)-($b_2$) with Figs. \ref{2bd_g0} ($b_1$)-($b_2$). 
As a case example, for $g_{BF}=-3$ it holds that $\rho^{(2)}_{FF}(x_1=4,x_2=4) \approx 0.1$ when $g_{BB} = 0$ 
while $\rho^{(2)}_{FF}(x_1=4,x_2=-4)\approx 0.12$ for $g_{BB} = 1$, see Fig.~\ref{2bd_g0}($b_1$) and 
Fig. ~\ref{2bd_g1}($b_1$) respectively.

\section{Quench Dynamics}\label{quench_dynamics}

Up to now we have discussed the ground state properties of the harmonically trapped particle imbalanced BF mixture 
with $N_F=6$ and $N_B=2$ for different impurity-medium interaction strengths ranging from attractive to repulsive values. 
Importantly, we have identified the presence of attractive induced interactions for the non-interacting impurities and 
analyzed the competition between the direct $s$-wave repulsive interactions with the induced ones. 
Also, in all cases we have quantified the back-action of the impurities to their fermionic environment. 

Below, we explore the corresponding non-equilibrium dynamics of the impurities and the Fermi sea. 
The mixture is prepared in its ground-state configuration, as already discussed in Sec.~\ref{Ground state}, with zero impurity-medium coupling strength. 
The dynamics is triggered by applying a quench of this coupling towards either the repulsive [Sec.~\ref{repulsive_quench}] or the 
attractive [Sec.~\ref{attractive quench}] regime of interactions~\cite{volosniev2015real,mistakidis2019effective}. 
Our main objective is to inspect the dynamical emergence of induced impurity-impurity correlations and the pattern formation of the fermionic environment as a result of the impurities motion. 
In the subsequent analysis we first study the dynamics of two non-interacting impurities and then contrast our findings to the case of two repulsively interacting ones.

\subsection{Quench to repulsive interactions}\label{repulsive_quench} 

We focus first on the correlated dynamics of the BF mixture induced by a quench from a vanishing to repulsive impurity-medium interactions. 
The emergent dynamics is firstly analyzed by employing the corresponding single-particle density evolution of the participating 
components [Sec.~\ref{density_evol_repul}] and then by inspecting their two-body density matrix [Sec.~\ref{two_body_evol_repul}] in 
the course of the evolution. 
These observables enable us to gain an overview of the dynamical evolution and importantly shed light on 
the existence of impurity-impurity and bath correlations respectively.

\subsubsection{Single-particle density evolution}\label{density_evol_repul}

\begin{figure}[ht]
\centering
\includegraphics[width=0.47\textwidth]{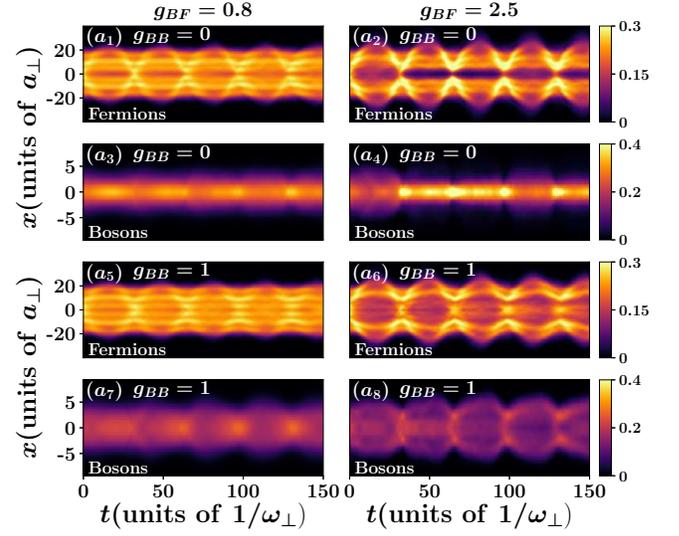}
\caption{One-body density evolution of [($a_1$), ($a_2$), ($a_5$), ($a_6$)] the Fermi sea $\rho^{(1)}_{F}(x,t)$ and [($a_3$), ($a_4$), ($a_7$), ($a_8$)] the bosonic impurities $\rho^{(1)}_{F}(x,t)$ upon considering an impurity-medium interaction quench from $g_{BF}=0$ to a finite repulsive value (see legend). 
The impurities are ($a_1$)-($a_4$) either free i.e. $g_{BB} = 0$ or ($a_5$)-($a_8$)  repulsively interacting with $g_{BB} = 1$. 
The system is confined in a harmonic trap with $\omega=1$ and comprises of $N_B = 2$ bosons immersed in Fermi sea of $N_F = 6$ fermions. 
It is initialized in its ground state with $g_{BF}=0$ and either $g_{BB}=0$ or $g_{BB}=1$.}  
\label{spd_dr}
\end{figure}

To gain an overview of the spatially resolved quench-induced dynamics of the BF mixture we show the corresponding single-particle density evolution of the impurities $\rho^{(1)}_{B}(x;t)$ and the Fermi sea $\rho^{(1)}_{F}(x;t)$ in Fig.~\ref{spd_dr} for different impurity-medium and impurity-impurity interaction strengths. 
Naturally, we commence our discussion on the system containing non-interacting impurities which provides the most clear signatures of induced correlations. 
Referring to weak post-quench interactions namely $g_{BF}=0.8$ the fermionic environment performs an overall 
breathing motion~\cite{huang2019breathing,boudjemaa2020breathing} manifested as a small amplitude periodic expansion and contraction of its cloud, see Fig.~\ref{spd_dr}($a_1$). 
The frequency of this global breathing mode is $\omega_F^{br}\approx 0.193\approx2\omega$ which is indeed in accordance to the corresponding theoretical predictions~\cite{bauch2009quantum,abraham2012quantum}. 
Moreover $\rho^{(1)}_{F}(x;t)$ exhibits at each time-instant of the evolution six in total shallow local maxima, namely three on the left and other three on the right side with respect to $x=0$, while a shallow density dip occurs around the trap center $x=0$. 
These local maxima are indicative of the six fermions present in the system and also the fact that majorly the first six single-particle eigenstates of the trap participate in the dynamics. 
On the other hand, the shallow dip of $\rho^{(1)}_{F}(x=0;t)$ is caused by the presence of the impurities at the same location. 
The impurities density $\rho^{(1)}_{B}(x;t)$ undergoes a very weak amplitude breathing dynamics [see Fig.~\ref{spd_dr}($a_3$)] characterized by a predominant 
frequency $\omega_{B}^{br}\approx 0.24$. 
Notice here that $\omega_B^{br}$ is slightly larger than $\omega_F^{br}$ since the impurities experience an effective potential, created by the external trap and the density of the Fermi sea, which possesses a larger than the trap frequency. 
Moreover, $\rho^{(1)}_{B}(x;t)$ completely overlaps with $\rho^{(1)}_{F}(x;t)$ throughout the time-evolution, thus indicating the miscible character of the dynamics. 
Recall that the two components are also miscible in the ground state of the system for $g_{BF}=0.8$ as discussed in Sec.~\ref{ground_state_density} and demonstrated in Figs. \ref{spd_g} ($a_1$),($b_1$). 

Turning to repulsively interacting impurities with $g_{BB}=1$ for the same quench amplitude i.e. from $g_{BF} = 0$ to $g_{BF} = 0.8$ we observe that a qualitatively similar to the 
above-described dynamics takes place when $g_{BB}=1$ for both the fermionic environment [Fig.~\ref{spd_dr}($a_5$)] and the 
bosons [Fig.~\ref{spd_dr}($a_7$)]. 
A notable difference is that $\rho^{(1)}_{B}(x;t)$ is broader in the $g_{BB}=1$ case due to the presence of the direct $s$-wave interaction, compare in particular Fig.~\ref{spd_dr}($a_3$) with Fig.~\ref{spd_dr}($a_7$). 
Note that this broadening of $\rho^{(1)}_{B}(x;t)$ for finite $g_{BB}$ occurs also in the ground state of the system, see Figs.~\ref{spd_g}($b_1$), ($b_2$). 
As expected, also the breathing amplitude of the impurities is larger when $g_{BB}=1$ while the frequency of this motion remains almost 
the same ($\omega_{B}^{br}\approx 0.22$) with the $g_{BB}=0$ case. 
This small deviation in the value of $\omega_{B}^{br}$ is attributed to interaction effects~\cite{schmitz2013quantum,kiehn2019spontaneous}. Consequently, due to the broader $\rho^{(1)}_{B}(x;t)$ when $g_{BB}=1$ the central dip in $\rho^{(1)}_{F}(x=0;t)$ occurring for $g_{BB}=0$ [Fig.~\ref{spd_dr}($a_1$)] becomes very shallow and almost disappears for $g_{BB}=1$ [Fig.~\ref{spd_dr}($a_3$)]. 
Otherwise, the inclusion of direct $s$-wave impurity-impurity interactions does not alter the characteristics of the system's dynamics at least on the single-particle level. 

Increasing the post-quench interaction strength e.g. to $g_{BF}=2.5$ gives rise to a much more intricate dynamics for both the non-interacting impurities [Fig.~\ref{spd_dr}($a_4$)] and the fermionic medium [Fig.~\ref{spd_dr}($a_2$)] when compared to the $g_{BF}=0.8$ quench amplitude. 
We remark that such a difference is already expected from the ground state properties of the system since for $g_{BF}=0.8$ the components are spatially overlapping (miscible) and become immiscible for $g_{BF}=2.5$, see also Figs. \ref{spd_g} ($a_1$), ($b_1$). 
In particular, the cloud of the Fermi sea exhibits a breathing oscillation with almost the same frequency $\omega_F^{br}\approx0.195$ 
as for the quench to $g_{BF}=0.8$. 
However, the amplitude of the contraction and expansion dynamics of $\rho^{(1)}_{F}(x;t)$ [Fig.~\ref{spd_dr}($a_2$)] is larger when compared to the smaller $g_{BF}=0.8$ [Fig.~\ref{spd_dr}($a_1$)] leading to a comparatively more excited medium in the former case. 
Accordingly, $\rho^{(1)}_{F}(x;t)$ appears to be in general wider for $g_{BF}=2.5$, a result that can again be traced back to the ground state density of the Fermi sea [Fig.~\ref{spd_g}($a_1$)]. 
Moreover, the local density humps building upon $\rho^{(1)}_{F}(x;t)$ [Fig.~\ref{spd_dr}($a_2$)] are found to be shallower (deeper) during the expansion (contraction) of the fermionic cloud for $g_{BF}=2.5$ than for $g_{BF}=0.8$. 
Importantly, the density dip of $\rho^{(1)}_{F}(x;t)$  around the trap center is substantially deeper when $g_{BF}=2.5$. 
This is a direct consequence of the emergent phase-separation being anticipated already from the ground state of the system for such strongly repulsive impurity-medium interactions, see also Figs.~\ref{spd_g} ($a_1$), ($b_1$). 

Of course, most of the above-described features of $\rho^{(1)}_{F}(x;t)$ are intimately connected with the corresponding behavior of the single-particle density of the bosons $\rho^{(1)}_{B}(x;t)$ [Fig.~\ref{spd_dr}($a_4$)] since the two components are inevitably interdependent due to their mutual finite coupling $g_{BF}$. 
Specifically, the impurities density $\rho^{(1)}_{B}(x;t)$ shows a relatively larger localization tendency [Fig.~\ref{spd_dr}($a_4$)] than for $g_{BF}=0.8$ [Fig.~\ref{spd_dr}($a_3$)] which is expected due to the aforementioned phase-separated behavior among the components. 
Moreover, $\rho^{(1)}_{B}(x;t)$ exhibits a weaker amplitude and 
larger frequency $\omega_{B}^{br}\approx 0.36$ breathing motion for $g_{BF}=2.5$ than for $g_{BF}=0.8$. 
The alteration of the impurities breathing frequency for $g_{BF}=2.5$ can in turn be explained within an effective potential picture. 
Indeed, as already argued in the ground state of the system the impurities can be viewed as trapped in the potential formed by 
the harmonic trap with a superimposed density of their Fermi sea. 
Since $\rho^{(1)}_{F}(x;t)$ is wider for increasing $g_{BF}$ also the impurities effective trapping frequency being r
elated to the breathing one is larger. 

The dynamics of interacting impurities with $g_{BB}=1$ following a quench to $g_{BF}=2.5$ as captured by  $\rho^{(1)}_{F}(x;t)$ 
[Fig.~\ref{spd_dr}($a_6$)] and $\rho^{(1)}_{B}(x;t)$ [Fig.~\ref{spd_dr}($a_8$)] is more involved than the $g_{BB}=0$ case, especially for long evolution times $t>40$. 
Evidently, the impurities exhibit a significantly broader density distribution for $g_{BB}=1$ [Fig.~\ref{spd_dr}($a_8$)] than 
for $g_{BB}=0$ [Fig.~\ref{spd_dr}($a_4$)], while performing a breathing motion of a larger amplitude and smaller 
frequency $\omega_{B}^{br}\approx 0.33$ in the former case. 
Furthermore, $\rho^{(1)}_{B}(x;t)$ initially ($t=0$) having a Gaussian profile deforms already within the initial stages of the dynamics ($t>5$) 
by developing three shallow humps being more pronounced during expansion and coming very close at the contraction points [Fig.~\ref{spd_dr}($a_8$)]. 
This behavior of $\rho^{(1)}_{B}(x;t)$ essentially indicates that for $t<5$ the interacting impurities dominantly occupy the lowest-lying 
single-particle eigenstate of their external potential while as time evolves the contribution of higher-lying eigenstates becomes significant 
and excitations are formed. 
This statement is also supported by the population of the individual bosonic orbitals $\phi_i^{B}(x,t)$ with $i=1,2,\dots,12$ 
(see also the discussion following Eq.~(\ref{1BD})) from which the first eight have a non-negligible population during the evolution (results not shown). 
The above-mentioned differences, regarding mainly the breathing mode and the structure of $\rho^{(1)}_{B}(x;t)$ for $g_{BB}=1$ compared to 
the $g_{BB}=0$ case are attributed to the presence of the direct $s$-wave repulsive interaction between the impurities. 
As expected, the features of $\rho^{(1)}_{B}(x;t)$ are also imprinted to a certain extent in the density of the fermionic environment 
$\rho^{(1)}_{F}(x;t)$ [Fig.~\ref{spd_dr}($a_6$)] due to the finite $g_{BF}$. 
Notably, $\rho^{(1)}_{F}(x;t)$ shows an arguably suppressed central dip compared to the $g_{BB}=0$ case due to 
the broader distribution of the impurities for $g_{BB}=1$. 
This in turn gives rise to an almost vanishing degree of phase-separation in the latter $g_{BB}=1$ case. 
Other properties, such as the amplitude and the frequency of the breathing mode remain almost the same as in the $g_{BB}=0$ scenario. 

Concluding this section, it is important to emphasize that the quench dynamics of non-interacting and interacting bosonic impurities 
differs noticeably already on the single-particle level, especially for large post-quench interaction strengths. 
This impact of the direct $s$-wave interaction of the impurities is also imprinted in the Fermi sea leading to changes in its 
pattern formation. 
As we shall explicate below, the origin of the above-mentioned differences is the presence of impurity-impurity induced interactions.

\subsubsection{Dynamics of impurity-impurity correlations}\label{two_body_evol_repul} 

\begin{figure}[ht]
\centering
\includegraphics[width=0.47\textwidth]{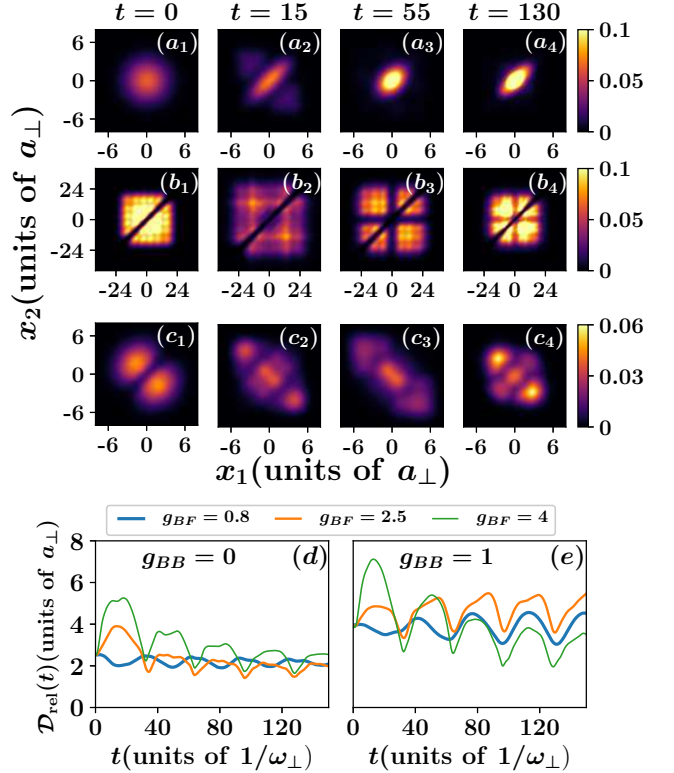}
\caption{Snapshots of the two-body reduced density of ($a_1$)-($a_4$) two non-interacting bosons $\rho^{(2)}_{BB}(x_1,x_2)$, ($b_1$)-($b_4$) two fermions of the medium $\rho^{(2)}_{FF}(x_1,x_2)$ and ($c_1$)-($c_4$) two repulsively interacting bosons $\rho^{(2)}_{BB}(x_1,x_2)$ at specific time-instants of the evolution (see legends). 
The system consists of $N_F=6$ fermions and $N_B=2$ bosons while it is confined in a harmonic trap with $\omega=0.1$. 
It is initialized in its ground state with $g_{BF}=0$ and either $g_{BB}=0$ or $g_{BB}=1$. 
To trigger the dynamics an interaction quench is performed from $g_{BF} = 0$ to $g_{BF} = 2.5$. 
Time-evolution of the relative distance $\mathcal{D}_{\rm rel}(t)$ between the two bosonic impurities with ($d$) $g_{BB} = 0$ and ($e$) $g_{BB} =1$ at distinct post-quench $g_{BF}$ values (see legend).}  
\label{2B_den_r}
\end{figure}

To track the spatially resolved dynamics of the two bosonic impurities with respect to one another we next invoke their 
two-particle density, $\rho^{(2)}_{BB}(x_1,x_2);t)$ [Eq.~(\ref{2BD})], which essentially provides the probability of measuring 
simultaneously one particle at position $x_1$ and the other at $x_2$. 
As a complementary measure of the impurities position we also calculate their relative distance $\mathcal{D}_{\rm rel}(t)$ [Eq.~(\ref{7})] during the time-evolution. 
This observable will allow us to identify whether the impurities interact among each other via induced correlations mediated by their host or 
they move independently~\cite{mistakidis2019correlated,mistakidis2020many}. 
Snapshots of $\rho^{(2)}_{BB}(x_1, x_2;t)$ at specific time-instants of the evolution upon considering a quench from $g_{BF}=0$ to $g_{BF}=2.5$ for the cases of $g_{BB}=0$ and $g_{BB}=1$ are presented in Figs.~\ref{2B_den_r}($a_1$)-($a_4$) and Figs.~\ref{2B_den_r}($c_1$)-($c_4$) respectively. 
Moreover, the corresponding $\mathcal{D}_{\rm rel}(t)$ when $g_{BB}=0$ [Fig.~\ref{2B_den_r}($d$)] and $g_{BB}=1$ [Fig.~\ref{2B_den_r}($e$)] is demonstrated for different post-quench interactions providing an overview of the impurity-impurity correlations. 

For non-interacting bosonic impurities, $\rho^{(2)}_{BB}(x_1,x_2;t=0)$ has a circular shape in the ($x_1,x_2$)-plane [Fig.~\ref{2B_den_r}($a_1$)] with a peak around $x_1,x_2\in [-2,2]$. 
Therefore, the bosons are likely to reside in this spatial region close to the trap center. 
However, in the course of the dynamics this shape of $\rho^{(2)}_{BB}(x_1,x_2;t)$ is drastically altered exhibiting an elongated diagonal and a suppressed anti-diagonal, see Figs.~\ref{2B_den_r}($a_2$)-($a_4$). 
Note that the anti-diagonal of the two-particle density of the impurities dictates their relative distance $\mathcal{D}_{\rm rel}(t)$ [Eq.~(\ref{7})]. 
The latter is illustrated in Fig.~\ref{2B_den_r}($d$) for a variety of post-quench $g_{BF}$ values. 
As it can be seen, in all cases $\mathcal{D}_{\rm rel}(t)$ undergoes a decaying amplitude oscillatory motion characterized by 
two dominantly participating frequencies which essentially correspond to the center-of-mass and relative coordinate 
breathing modes~\cite{schmitz2013quantum}, e.g. $\omega_1\approx 0.19$, $\omega_2\approx 0.24$ for $g_{BF}=0.8$ and 
$\omega_1\approx 0.19$, $\omega_2\approx 0.36$ when $g_{BF}=2.5$. 
Indeed, the oscillatory behavior of $\mathcal{D}_{\rm rel}(t)$ reflects the breathing motion of the impurities cloud already identified in 
the dynamics of $\rho^{(1)}_{B}(x;t)$ [Fig.~\ref{spd_dr}($a_3$)]. 
This is also imprinted in the modulated shape of $\rho^{(2)}_{BB}(x_1,x_2;t)$ [Figs.~\ref{2B_den_r}($a_2$-($a_4$)], e.g. 
the anti-diagonal of $\rho^{(2)}_{BB}(x_1,x_2;t)$ is more expanded at $t=55$ compared to $t=130$. 
Also, the oscillation amplitude of $\mathcal{D}_{\rm rel}(t)$ and as a consequence of the breathing mode is enhanced 
for a larger post-quench $g_{BF}$, see also Figs.~\ref{spd_dr}($a_3$), ($a_7$). 
Importantly the decaying amplitude in time of $\mathcal{D}_{\rm rel}(t)$, and thus the elongated shape 
of $\rho^{(2)}_{BB}(x_1,x_2;t)$ across its diagonal, implies that the impurities tend to approach 
each other during the dynamics and since they are non-interacting ($g_{BB}=0$) they experience an effective 
attraction mediated by the fermionic environment. 

Proceeding we examine the role played by the direct $s$-wave repulsive contact interaction between the impurities and its competition with the induced interactions on the quench dynamics. 
Due to the finite impurity-impurity repulsion, herein $g_{BB}=1$, the bosons initially ($t=0$) reside one in the left ($x<0$) and the other in the right ($x>0$) side of the trap, see the pronounced anti-diagonal distribution of $\rho^{(2)}_{BB}(x_1,x_2=-x_1;t=0)$ in Fig.~\ref{2B_den_r}($c_1$). 
In contrast, after the quench ($t>0$) three distinct segments develop in $\rho^{(2)}_{BB}(x_1,x_2;t)$ [Figs.~\ref{2B_den_r}($c_2$)-($c_4$)]. 
Namely the impurities are either very close to each other around the trap center [see the diagonal of $\rho^{(2)}_{BB}(x_1,x_2;t)$] or they remain spatially separated with one of them located in the left and the other in the right side of the trap with respect to $x=0$ [see the anti-diagonal of $\rho^{(2)}_{BB}(x_1,x_2;t)$]. 
This two-body superposition is a consequence of the competition between the direct repulsive and induced 
attractive interactions~\cite{mistakidis2020many,mistakidis2020induced}. 
Inspecting now $\mathcal{D}_{\rm rel}(t)$ for different post-quench values of $g_{BF}$ [Fig.~\ref{2B_den_r}($d$)] we can 
readily see that it performs oscillations possessing two predominant frequencies, for instance 
$\omega_1\approx 0.22$, $\omega_2\approx 0.19$ when $g_{BF}=0.8$ and $\omega_1\approx 0.33$, $\omega_2\approx 0.19$ if $g_{BF}=2.5$. 
These frequencies are again related to the center-of-mass and relative coordinate breathing modes respectively. 
Interestingly, the oscillation amplitude of $\mathcal{D}_{\rm rel}(t)$ e.g. for $g_{BF}=0.8$ and $g_{BF}=2.5$ is almost constant while for $g_{BF}=4$ it shows a decaying tendency. 
This means that in the latter case the induced attraction tends to surpass the impurities direct repulsion. 
Finally, we remark that the oscillation amplitude (decay rate) of $\mathcal{D}_{\rm rel}(t)$ for fixed $g_{BF}$ is in general larger (smaller) when $g_{BB}=1$ [Fig.~\ref{2B_den_r}($e$)] compared to $g_{BB}=0$ [Fig.~\ref{2B_den_r}($d$)], thus evincing the presence of the direct impurities repulsion.

\subsubsection{Correlations of the fermionic environment}\label{two_body_evol_repul_bath} 

To complement our study we then investigate the correlation patterns of the Fermi sea as encoded in its two-body 
density $\rho^{(2)}_{FF}(x_1,x_2;t)$ shown in Figs.~\ref{2B_den_r}($b_1$)-($b_4$) at specific time-instants after a 
quench from $g_{BF}=0$ to $g_{BF}=2.5$ for $g_{BB}=0$. 
A correlation hole occurs along the diagonal of $\rho^{(2)}_{FF}(x_1,x_2;t)$ throughout the evolution due to the Pauli exclusion principle. 
Also, an expansion [Fig.~\ref{2B_den_r}($b_2$)] and contraction [Fig.~\ref{2B_den_r}($b_4$)] of the anti-diagonal 
of $\rho^{(2)}_{FF}(x_1,x_2;t)$ 
takes place which manifest the collective breathing motion of the fermionic cloud~\cite{kwasniok2020correlated}, see 
also Fig.~\ref{spd_dr}($a_2$). 
Moreover, a depletion along the $x_1=0$ and $x_2=0$ spatial regions is observed indicating that it is more likely for 
one fermion to be located in the vicinity of a density hump at $x<0$ and the other one being symmetrically placed 
with respect to the trap center, see also Fig.~\ref{spd_dr}($a_2$).

\subsection{Quench to attractive interactions}\label{attractive quench} 

In the following, we shall study the dynamical response of the impurities and the fermionic bath after a quench from $g_{BF}=0$ to 
the attractive ($g_{BF}<0$) impurity-medium interaction regime. 
To quantify the arising distinctive dynamical features we analyze the single-particle density [Sec.~\ref{density_evol_attract}] 
and the corresponding two-body density [Sec.~\ref{two_body_evol_attract}] evolution of the participating components. 
As in the previous section, we first discuss the time-evolution of two non-interacting ($g_{BB} = 0$) impurities and subsequently 
compare to the case of two repulsively interacting ($g_{BB} = 1.0$) ones.

\subsubsection{Density evolution}\label{density_evol_attract}

\begin{figure}[ht]
\centering
\includegraphics[width=0.47\textwidth]{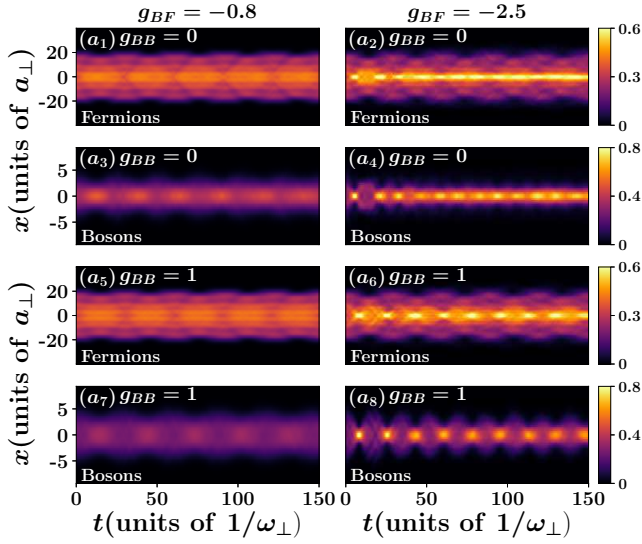}
\caption{Time-evolution of the one-body density of [($a_1$), ($a_2$), ($a_5$), ($a_6$)] the fermionic medium $\rho^{(1)}_{F}(x,t)$ and [($a_3$), ($a_4$), ($a_7$), ($a_8$)] the bosonic impurities $\rho^{(1)}_{B}(x,t)$ after an interaction quench of the boson-fermion coupling constant from $g_{BF}=0$ to different attractive values (see legend). 
The impurities are considered to be ($a_1$)-($a_4$) free i.e. $g_{BB} = 0$ and ($a_5$)-($a_8$) repulsively interacting with $g_{BB} = 1$. 
The harmonically trapped mixture with $\omega=1$ consists of $N_F = 6$ fermions and $N_B = 2$ bosons while it is prepared in its ground state with $g_{BF}=0$ and either $g_{BB}=0$ or $g_{BB}=1$.} 
\label{s_den_a}
\end{figure}

The spatio-temporal evolution of $\rho^{(1)}_{F}(x;t)$ and $\rho^{(1)}_{B}(x;t)$ after a quench from $g_{BF}=0$ to $g_{BF}=-0.8$ 
for $g_{BB}=0$ is presented in Fig.~\ref{s_den_a} ($a_1$) and Fig.~\ref{s_den_a} ($a_3$) respectively. 
As a consequence of the interaction quench both the impurities and the fermionic clouds undergo a collective weak amplitude 
breathing motion identified by their contraction and expansion dynamics~\cite{huang2019breathing,mistakidis2019correlated}. 
The breathing frequency of the fermionic bath is $\omega_F^{br}\approx 0.208\approx2\omega$ while for the impurities it corresponds to 
$\omega_B^{br} \approx 0.251$ since they experience a modified external potential composed of the harmonic oscillator and the density of their host, see for details Refs.~\cite{mistakidis2020many,kiehn2019spontaneous}. 
Importantly, the attractive impurity-medium coupling results in the formation of a shallow density hump in $\rho^{(1)}_{F}(x;t)$ [Fig.~\ref{s_den_a}($a_1$)] at the instantaneous location of the impurities [Fig.~\ref{s_den_a}($a_3$)] i.e. around the trap center. 
Turning to repulsively interacting impurities where $g_{BB}=1$ we observe that a similar to the above-described dynamics takes place, see Figs.~\ref{s_den_a} ($a_5$), ($a_7$). 
However, the expansion amplitude of $\rho^{(1)}_{B}(x;t)$ is larger and the breathing frequency $\omega_B^{br}\approx 0.226$ is slightly smaller compared to the $g_{BB}=0$ scenario due to the inclusion of direct $s$-wave repulsive interactions. 
Also, since $\rho^{(1)}_{B}(x;t)$ is relatively wider than for $g_{BB}=0$ the density hump developed in $\rho^{(1)}_{F}(x;t)$ in 
the latter case almost disappears for $g_{BB}=1$.   

Following a quench to stronger impurity-medium interactions, e.g. $g_{BF}=-2.5$, leads to a more intricate response of both components than for $g_{BF}=-0.8$, see Figs.~\ref{s_den_a}($a_2$), ($a_4$), ($a_6$), ($a_8$). 
Referring to the system containing non-interacting impurities [Figs.~\ref{s_den_a}($a_2$), ($a_4$)] we observe that $\rho^{(1)}_{B}(x;t)$ has a pronounced spatial localization tendency around the trap center while performing an ``irregular'' weak amplitude breathing dynamics. 
The latter is characterized by two dominant frequencies, namely $\omega_{B_1}^{br}=0.234$ and $\omega_{B_2}^{br}=0.263$ corresponding to the 
center-of-mass and relative coordinate breathing mode respectively. 
Since these frequencies are close the dynamics of $\rho^{(1)}_{B}(x;t)$ is reminiscent of a beating pattern. 
We remark that a similar time-evolution takes place also for bosonic impurities immersed in a bosonic background~\cite{mistakidis2020many}.
Accordingly, as a result of this sharply peaked distribution of $\rho^{(1)}_{B}(x;t)$ in the vicinity of $x=0$ there is an accumulation 
of the fermionic density at the same location due to the finite $g_{BF}$. 
Indeed, a prominent density hump builds upon $\rho^{(1)}_{F}(x;t)$ [Fig.~\ref{s_den_a}($a_2$)] which otherwise exhibits collective 
breathing oscillations of a frequency $\omega_F^{br}\approx 0.211$. 
The dynamical response is somewhat changed when considering interacting impurities ($g_{BB}=1$) as depicted in Figs.~\ref{s_den_a}($a_6$), ($a_8$). 
The impurities possess a comparatively wider density distribution than for $g_{BB}=0$ as a consequence of their finite repulsion, $g_{BB}=1$. 
Also, the amplitude of their breathing motion is slightly larger compared to the one of $g_{BB}=0$ and the participating frequencies 
$\omega_{B_1}^{br}=0.355$ and $\omega_{B_2}^{br}=0.376$ are very close thereby producing a beating pattern (hardly visible 
in Fig.~\ref{s_den_a}($a_8$)) manifested by the periodic 
appearance of sharp peaks in $\rho^{(1)}_{B}(x;t)$ as a result of its contraction. 
The differences in the dynamics of non-interacting and interacting impurities for fixed post-quench $g_{BF}$ will be further analyzed in the next section. 
This beating pattern is also imprinted in $\rho^{(1)}_{F}(x;t)$ which as a back-action develops humps which follow the location of the impurities. 
Otherwise, $\rho^{(1)}_{F}(x;t)$ performs a breathing mode of almost the same amplitude and equal 
frequency $\omega_{F}^{br}\approx0.2$ compared to the $g_{BB}=0$ case.

\subsubsection{Evolution of impurity-impurity correlations}\label{two_body_evol_attract} 

\begin{figure}[ht]
\centering
\includegraphics[width=0.47\textwidth]{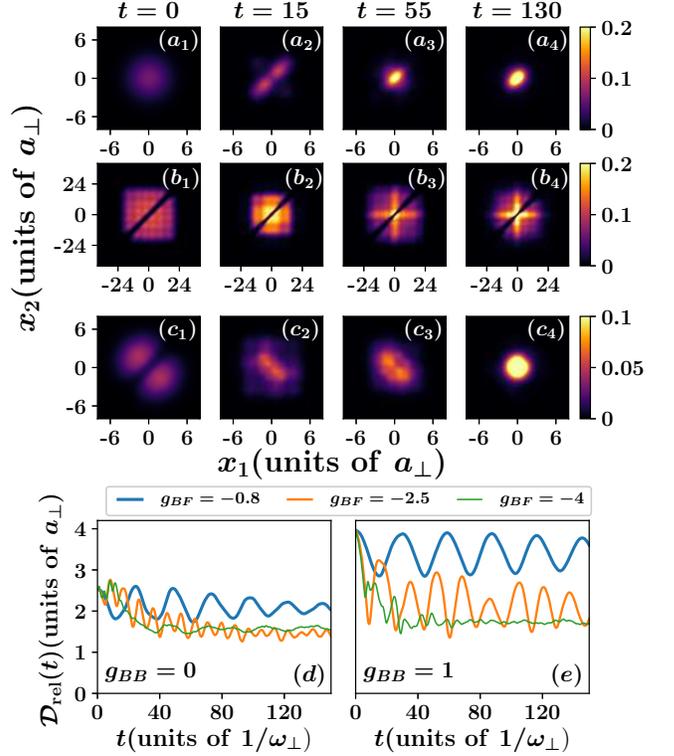} 
\caption{Instantaneous profiles of the two-body reduced density of ($a_1$)-($a_4$) two non-interacting bosons $\rho^{(2)}_{BB}(x_1,x_2)$, ($b_1$)-($b_4$) two fermions of the bath $\rho^{(2)}_{FF}(x_1,x_2)$ and ($c_1$)-($c_4$) two repulsively interacting bosons $\rho^{(2)}_{BB}(x_1,x_2)$. 
The BF mixture contains $N_F = 6$ fermions and $N_B =  2$ bosons. 
It is confined in a harmonic trap with $\omega = 0.1$ and it is prepared in its ground state with $g_{BF} = 0$ and either $g_{BB}=0$ or $g_{BB}=1$. 
The dynamics is triggered upon considering an impurity-medium interaction quench from $g_{BF} = 0$ to $g_{BF} = -2.5$. 
Temporal-evolution of the boson-boson relative distance $\mathcal{D}_{\rm rel}(t)$ for ($d$) $g_{BB} = 0$ and ($e$) $g_{BB} =1$ at specific post-quench $g_{BF}$ couplings (see legend).}  
\label{2B_den_a}
\end{figure}

To identify and consequently characterize the nature of the impurity-impurity correlations in the course of the dynamics we monitor the two-body density $\rho^{(2)}_{BB}(x_1,x_2;t)$ and relative distance $\mathcal{D}_{\rm rel}(t)$ of the impurities depicted in Fig.~\ref{2B_den_a} for different quench amplitudes. 
Figures~\ref{2B_den_a}($a_1$)-($a_4$) show snapshots of $\rho^{(2)}_{BB}(x_1,x_2;t)$ upon considering a quench of two non-interacting impurities from $g_{BF}=0$ to $g_{BF}=-2.5$. 
Initially, $t=0$, the impurities lie in the vicinity of the trap center since $\rho^{(2)}_{BB}(x_1,x_2;t)$ is non-zero within the spatial region $x_1,x_2\in [-2,2]$ [Fig.~\ref{2B_den_a}($a_1$)]. However, as time evolves, the two bosons start to occupy a smaller spatial region and therefore approach each other, a tendency that becomes evident by the gradual shrinking of $\rho^{(2)}_{BB}(x_1,x_2;t)$ across its diagonal accompanied by the depression of its anti-diagonal, see Figs.~\ref{2B_den_a}($a_2$)-($a_4$). 

As it has already been discussed in Sec.~\ref{two_body_evol_repul} the shape of the anti-diagonal of $\rho^{(2)}_{BB}(x_1,x_2;t)$ is well captured by $\mathcal{D}_{\rm rel}(t)$ which is presented in Fig.~\ref{2B_den_a}($d$) for distinct post-quench values of $g_{BF}$. 
Evidently $\mathcal{D}_{\rm rel}(t)$ oscillates irrespectively of $g_{BF}$, a behavior that corresponds to the breathing motion of the impurities and can also be inferred from the weak expansion [Fig.~\ref{2B_den_a}($a_3$)] and contraction [Fig.~\ref{2B_den_a}($a_2$)] of $\rho^{(2)}_{BB}(x_1,x_2=-x_1;t)$. 
Its evolution contains a multitude of frequencies whose number and value depend on $g_{BF}$ and refer to the underlying 
breathing motion, e.g. for $g_{BF}=-2.5$ the dominantly involved frequencies are $\omega_1=0.234$ and $\omega_2=0.263$ respectively. 
Also, the oscillation amplitude of $\mathcal{D}_{\rm rel}(t)$ is smaller for a larger $\abs{g_{BF}}$ which is in accordance to the localization tendency of the impurities for quenches to stronger impurity-medium attractions. 
Importantly, $\mathcal{D}_{\rm rel}(t)$ exhibits a decaying tendency in time which is more pronounced for increasing $\abs{g_{BF}}$ and shows a saturation behavior for quite strong attractions, e.g. $g_{BF}=-4$ here, and long evolution times $t>80$. 
This latter decaying behavior is again a manifestation of the presence of attractive induced impurity-impurity interactions. 
It is also worth commenting at this point that the suppression of the anti-diagonal of $\rho^{(2)}_{BB}(x_1,x_2;t)$ or equivalently 
the decay of $\mathcal{D}_{\rm rel}(t)$ is significantly more pronounced for quenches towards the attractive interaction regime than 
in the repulsive one, e.g. compare Fig.~\ref{2B_den_r}($d$) and Fig.~\ref{2B_den_a}($d$). 
Therefore, we can infer the generation of stronger attractive induced interactions for quenches in the attractive than in the repulsive 
impurity-medium interaction regime~\cite{mistakidis2020induced}, a result that also holds for the ground state of the system as 
explicated in Sec.~\ref{corel_ground}. 

On the other hand, the two-body dynamics of two repulsively interacting impurities, here $g_{BB}=1$, subjected to a quench 
from $g_{BF}=0$ to $g_{BF}=-2.5$ showcases quite different characteristics from the $g_{BB}=0$ case, compare in particular 
Figs.~\ref{2B_den_a}($c_1$)-($c_4$) with Figs.~\ref{2B_den_r}($c_1$)-($c_4$). 
Indeed, even for the ground state of the system ($t=0$) the impurities, since $g_{BB}$ is finite, are spatially separated with the one residing 
at $x<0$ and the other at $x>0$ as can be inferred from the correlation hole of $\rho^{(2)}_{BB}(x_1,x_2=x_1;t=0)$ in Fig.~\ref{2B_den_a}($c_1$). 
After the quench, they oscillate between two distinct configurations. 
Namely they either stay separated, see the elongated anti-diagonal of $\rho^{(2)}_{BB}(x_1,x_2;t)$ in Figs.~\ref{2B_den_a}($c_2$), ($c_3$), or they move close to each other, see their bunching tendency in the region $x_1,x_2\in [-2,2]$ in Fig.~\ref{2B_den_a}($c_4$). 
This behavior is caused by the competition of their inherent repulsive contact interaction and the induced attraction mediated by the fermionic environment~\cite{huber2019medium}. 

To understand better the aforementioned competing mechanism we present the time-evolution of the impurities relative distance $\mathcal{D}_{\rm rel}(t)$ for a variety of post-quench interactions $g_{BF}$ in Fig.~\ref{2B_den_a}($e$). 
As it can be directly seen, the response of $\mathcal{D}_{\rm rel}(t)$ depends crucially on $g_{BF}$. Indeed for quenches to weak attractions, e.g. $g_{BF}=-0.8$, $\mathcal{D}_{\rm rel}(t)$ oscillates with an almost constant amplitude and a 
dominant frequency $\omega_1=0.226$. 
However, by increasing the quench amplitude e.g. to $g_{BF}=-2.5$ $\mathcal{D}_{\rm rel}(t)$ performs ``irregular'' oscillations characterized by multiple frequencies and importantly a decaying amplitude. 
This decay is more pronounced for larger attractions e.g. $g_{BF}=-4$ where $\mathcal{D}_{\rm rel}(t)$ drops at the early stages of the dynamics and saturates to a fixed value for $t>50$. 
As a consequence, we can deduce that the attractive induced interactions become stronger for quenches towards larger impurity-medium attractions and gradually dominate with respect to the impurities direct repulsion. 
Note here that such a mechanism is also present for quenches to repulsive impurity-medium interactions, see Fig.~\ref{2B_den_r}($e$), but it is apparently less effective compared to the attractive $g_{BF}$ quench scenario. 
\begin{figure*}[ht]
\centering
\includegraphics[width=\textwidth]{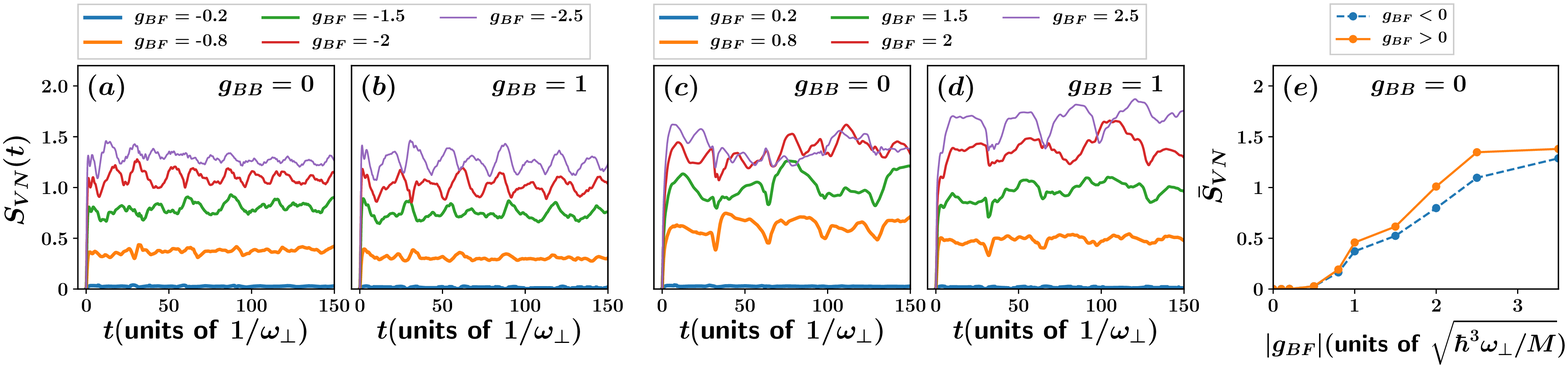}
\caption{Temporal-evolution of the Von-Neumann entropy $S_{VN}(t)$ for different post-quench ($a$)-($b$) attractive and ($c$)-($d$) repulsive impurity-medium interaction strengths (see legends). 
The bosonic impurities are considered to be either [($a$), ($c$)] non-interacting $g_{BB}=0$ or [($b$), ($d$)] interacting with $g_{BB}=1$. 
($e$) Time-averaged Von-Neumann entropy $\bar{S}_{VN}$ for distinct post-quench $\abs{g_{BF}}$ attractive and repulsive values (see legend) when $g_{BB}=0$. 
The solid and dashed lines provide a guide to the eye. 
The BF mixture consists of $N_B=2$ bosons and $N_F=6$ fermions with equal masses and are confined in the same harmonic trap with $\omega = 0.1$.}  
\label{fig:vn_ent}
\end{figure*}

\subsubsection{Correlations of the Fermi bath}\label{two_body_evol_bath_attract}

Turning to the Fermi sea we observe the appearance of completely different correlation patterns building upon $\rho^{(2)}_{FF}(x_1,x_2;t)$ when compared to the repulsive quench scenario [Figs.~\ref{2B_den_r}($c_1$)-($c_4$)] as demonstrated in Figs.~\ref{2B_den_a}($c_1$)-($c_4$) for a quench to $g_{BF}=-2.5$ in the system with two non-interacting ($g_{BB}=0$) impurities. 
As expected, a correlation hole exists for the entire time-evolution due to the fermionic character of the bath~\cite{erdmann2019phase,kwasniok2020correlated}. 
Initially, $t=0$, the fermions are symmetrically placed with respect to $x=0$ and predominantly reside one at $x>0$ and the other 
at $x<0$, see the bright spots close to the diagonal in Fig.~\ref{2B_den_a}($c_1$). 
Following the quench a cross-like correlation pattern appears in $\rho^{(2)}_{BB}(x_1,x_2;t)$ which becomes more elongated across the spatial regions lying in the vicinity of $x_1=0$ and $x_2=0$ [Figs.~\ref{2B_den_a}($c_3$), ($c_4$)]. 
This cross-like correlation pattern is the two-body analogue of the accumulation of the Fermi density $\rho^{(1)}_{F}(x;t)$ [Fig.~\ref{spd_dr}($a_2$)] around the trap center and more precisely in the vicinity of the position of the impurities due to the attractive $g_{BF}$. 
Thus it is a direct imprint of the impurities motion into their host, see also the elongated diagonal of $\rho^{(2)}_{BB}(x_1,x_2;t)$ in Figs.~\ref{2B_den_a}($a_2$)-($a_4$), evincing that for strong attractive $g_{BF}$ the fermions move close to the trap center, a mechanism that competes with their inherent Fermi pressure.

\subsection{Entanglement dynamics}\label{entanglemet_dynamics} 

To further unveil the degree of impurity-medium correlations during the quench dynamics of the BF mixture we employ the time-evolution of the Von-Neumann entropy $S_{VN}(t)$ [Eq.~\eqref{VN}]. 
This quantity provides a measure of the overall build up of the impurity-medium entanglement~\cite{mukherjee2020pulse,theel2020entanglement,kwasniok2020correlated} and also reveals the complexity of the time-evolved post-quench state of the system. 
The dynamics of $S_{VN}(t)$ after a quench to attractive (repulsive) interactions for the system of two non-interacting and interacting impurities is shown in Figs.~\ref{fig:vn_ent}($a$) and ($b$) [Figs.~\ref{fig:vn_ent}($c$) and ($d$)] respectively. 

Focusing on the attractive post-quench interaction regime [Figs.~\ref{fig:vn_ent}($a$), ($b$)] we observe that independently of the inclusion 
of direct $s$-wave impurity-impurity interactions $S_{VN}(t=0)=0$ and hence the components are initially non-entangled. 
However, directly after the quench an appreciable impurity-medium entanglement generation takes place in all cases 
since $S_{VN}(t)\neq 0$~\cite{mistakidis2019correlated,mistakidis2019quench,mukherjee2020pulse}. 
More precisely, an almost ballistic linear growth of $S_{VN}(t)$ is manifested at the very early stages of the dynamics ($t<5$) accompanied by a fluctuating behavior of $S_{VN}(t)$ around a fixed $g_{BF}$-dependent value at later evolution times. 
Notice that for both $g_{BB}=0$ and $g_{BB}=1$ the response of $S_{VN}(t)$ shows a hierarchy in terms of $g_{BF}$, namely it acquires 
larger values for stronger attractions. 
Also, the temporal fluctuations of $S_{VN}(t)$ deep in the evolution are suppressed for quenches to weak attractions, e.g. compare $S_{VN}(t)$ for $g_{BF}=-0.8$ and $g_{BF}=-2.5$ in Figs.~\ref{fig:vn_ent}($a$), ($b$). 
The latter means that for larger post-quench attractions the system is in a more complicated many-body superposition involving a 
larger amount of states [see also Eq.~(\ref{4})] than for smaller negative $g_{BF}$ values. 
This situation holds equal for fixed $g_{BF}$ but increasing $g_{BB}$. 
Indeed, it becomes apparent by inspecting $S_{VN}(t)$ for fixed $g_{BF}$ between the $g_{BB}=0$ and $g_{BB}=1$ cases that in the latter case the temporal fluctuations of $S_{VN}(t)$ are enhanced, especially for a larger $\abs{g_{BF}}$. 
We remark that the saturating tendency of $S_{VN}(t)$ for long times can be attributed to the finite size of the system~\cite{Calabrese_2005}, i.e. if the system would have been infinite then $S_{VN}(t)$ should increase linearly in time throughout the time-evolution. 

A similar to the above-described phenomenology regarding the entanglement dynamics takes place also during the unitary evolution of the system for quenches towards the repulsive impurity-medium interaction regime for both non-interacting [Fig.~\ref{fig:vn_ent}($c$)] and interacting [Fig.~\ref{fig:vn_ent}($d$)] impurities. 
Indeed, the sudden increase of $g_{BF}$ leads to entanglement formation since $S_{VN}(t)\neq 0$ while $S_{VN}(t=0)=0$. 
As for $g_{BF}<0$, here also $S_{VN}(t)$ increases linearly for $t<5$ and subsequently oscillates around a mean value, see Figs.~\ref{fig:vn_ent}($c$), ($d$). 
Interestingly, the degree of entanglement is larger for quenches to the repulsive than the attractive interaction regimes, e.g. compare Fig.~\ref{fig:vn_ent}($a$) with Fig.~\ref{fig:vn_ent}($c$). 
This fact evinces that a larger amount of dynamical impurity-medium entanglement is established in the repulsive interaction regime. 
To support our argument we exemplarily showcase the time-averaged Von-Neumann entropy, defined as $\bar{S}_{VN} = (1/T)\int_{0}^{T}dtS_{VN}(t)$ with $T$ being the considered evolution time, in Fig.~\ref{fig:vn_ent}($e$) for varying post-quench repulsive ($g_{BF}>0$) and attractive ($g_{BF}<0$) impurity-medium interactions in the system containing the non-interacting impurities. 
As it can be readily seen, irrespectively of the quench direction $\bar{S}_{VN}$ increases monotonously with increasing magnitude of $g_{BF}$. 
However, it is also apparent that $\bar{S}_{VN}$ is in general slightly larger for quenches to repulsive than to attractive interactions at a specific post-quench $\abs{g_{BF}}$.

\section{Conclusions}\label{conclusion} 

We have unraveled the role of induced correlations and pattern formation in the ground state and the non-equilibrium quantum dynamics of two bosonic impurities embedded in a fermionic environment. 
The one-dimensional Bose-Fermi mixture is harmonically trapped and the time-evolution is initiated upon considering a quench of the 
impurity-medium coupling from a vanishing towards the repulsive or the attractive interaction regime. 
Inspecting both one- and two-body observables enables us to expose correlation-induced phenomena mediated by the host, analyze the 
competition of induced interactions and direct $s$-wave ones, the emergent phase-separation processes and the underlying entanglement dynamics. 

Referring to the ground state of two non-interacting bosonic impurities it is shown that on the single-particle level they phase-separate 
with the Fermi sea for strong repulsions and accumulate at the trap center together with their environment for large attractions, otherwise they are miscible. 
In the system of two repulsively interacting impurities the boundaries of the aforementioned regions are shifted to larger interactions. 
Importantly, we identify the presence of induced impurity-impurity interactions mediated by the fermionic environment, in the system with non-interacting bosons, for either increasing impurity-medium repulsion or attraction. 
For repulsively interacting impurities we elaborate on the competition of induced and direct interactions with the latter (former) dominating for 
repulsive (attractive) impurity-medium couplings, evincing that the strength of induced interactions is larger for attractive impurity-bath interactions. 
Inspecting the two-body correlation function of the Fermi sea we showcase that two fermions are likely to remain far apart (approach each other) for larger impurity-medium repulsions (attractions). 

We trigger the dynamics by suddenly changing the impurity-medium interaction strength from zero to finite repulsive or attractive values. 
A quench to repulsive interactions induces in both components a collective breathing motion. 
The impurities breathing frequency and amplitude depend on the post-quench coupling and their interacting nature. 
Moreover, a dynamical phase-separation occurs for quenches to large repulsions with the impurities residing at the origin and the 
fermionic environment splitting into two symmetric density branches with respect to the trap center. 
Here, two fermions are likely to lie one on the left and the other on the right density branch. 
Interestingly, induced impurity-impurity correlations mediated by the host are manifested in the course of the evolution of two 
non-interacting impurities and become more pronounced for quenches to stronger repulsions. 
On the other hand, monitoring the dynamics of repulsively interacting impurities we showcase the competition of induced and 
direct interactions with the latter prevailing and enforcing the impurities to be in a two-body superposition. 
The impact of induced interactions is also captured by the decaying amplitude in time of the impurities relative distance, which is 
clearly more prominent for non-interacting bosonic impurities. 

For quenches to attractive impurity-medium interactions both components perform an overall breathing motion, whose amplitude and frequency regarding the impurities are impacted by the considered impurity-impurity and post-quench impurity-medium couplings. 
Remarkably, a beating pattern appears on the single-particle level stemming from the involvement of two nearly resonant breathing frequencies in the dynamics of the impurities due to the dominant nature of their attractive induced interactions. 
Furthermore, the impurities exhibit a spatial localization tendency around the trap center causing a density accumulation of the Fermi sea at their instantaneous location. 
This mechanism becomes more pronounced for quenches to larger attractions and it is imprinted as a cross-like correlation pattern in the Fermi sea and dictates the dominant presence of attractive induced interactions whose strength is enhanced for quenches to larger attractions. 
Indeed they can even gradually surpass the direct impurity-impurity repulsive coupling, a result that is also evident by the prominent decaying amplitude of the impurities relative distance during the time-evolution. 

Moreover, by measuring the Von-Neumann entropy we explicate that in all cases the impurity-medium entanglement rises in a linear manner at the initial stages of the dynamics and afterwards it exhibits a fluctuating 
behavior around a constant value. 
Also, the entanglement exhibits a hierarchy by means that it is larger for fixed impurity (post-quench impurity-medium) interaction 
and increasing quench amplitude (impurity coupling). 

There are several research directions that can be pursued in future endeavors. 
An intriguing perspective is to study the robustness of the discussed phenomena in the presence of finite 
temperature effects~\cite{tajima2019thermal} and in particular their impact on the impurities induced interactions. 
Certainly, the generalization of our findings to higher dimensional settings as well as to larger impurity concentrations is desirable. 
Another interesting direction would be to consider an additional long-range interparticle interaction potential~\cite{kain2014polarons} 
and unravel the corresponding quench induced dynamics. 
The emergent quasiparticle properties~\cite{schmidt2018universal,mistakidis2019repulsive} such as the lifetime, residue, 
effective mass and induced interactions are of particular interest.


\begin{acknowledgments} 
K.M. acknowledges a research fellowship (Funding ID no 57381333) from the Deutscher Akademischer Austauschdienst (DAAD). 
S. I. M. gratefully acknowledges financial support in the framework of the Lenz-Ising Award of the University of  Hamburg. 
P.S. is grateful for financial support by the Deutsche Forschungsgemeinschaft (DFG) in the framework 
of the SFB 925 ``Light induced dynamics and control of correlated quantum systems''. 
\end{acknowledgments}

\appendix

\section{Details of the many-body simulations and their convergence}\label{convergence}

In this appendix we provide a brief overview of the deployed variational approach, i.e. the multi-layer multi-configurational time-dependent Hatree method for atomic mixtures (ML-MCTDHX) utilized in the main text 
and also argue about the convergence of our results. 
ML-MCTDHX is an \textit{ab-initio} approach for solving the time-dependent MB Schr{\"o}dinger equation of multicomponent ultracold atom 
systems consisting of bosonic~\cite{mistakidis2018correlation,mistakidis2020many,mistakidis2020pump} or 
fermionic~\cite{erdmann2019phase,kwasniok2020correlated} species which could also possess spin degrees-of-freedom~\cite{koutentakis2020interplay,mittal2020many}. 
An important facet of this method is the expansion of the system's MB wavefunction in terms of a time-dependent and variationaly optimized basis set, see Eqs.~(\ref{4}) and (\ref{5}). 
Such a treatment is tailored to consider the basis states that are energetically favorable at each time-instant of the evolution and as a result to span the relevant subspace of the Hilbert space more efficiently than when using methods which rely on a time-independent basis. 
Consequently, it is possible to capture all relevant intra- and intercomponent correlations according to the system and driving protocol under investigation as well as to address setups with mesoscopic particle numbers~\cite{katsimiga2017dark,mistakidis2019correlated,mistakidis2018correlation}. 
\begin{figure}[ht]
\centering
\includegraphics[width=0.45\textwidth]{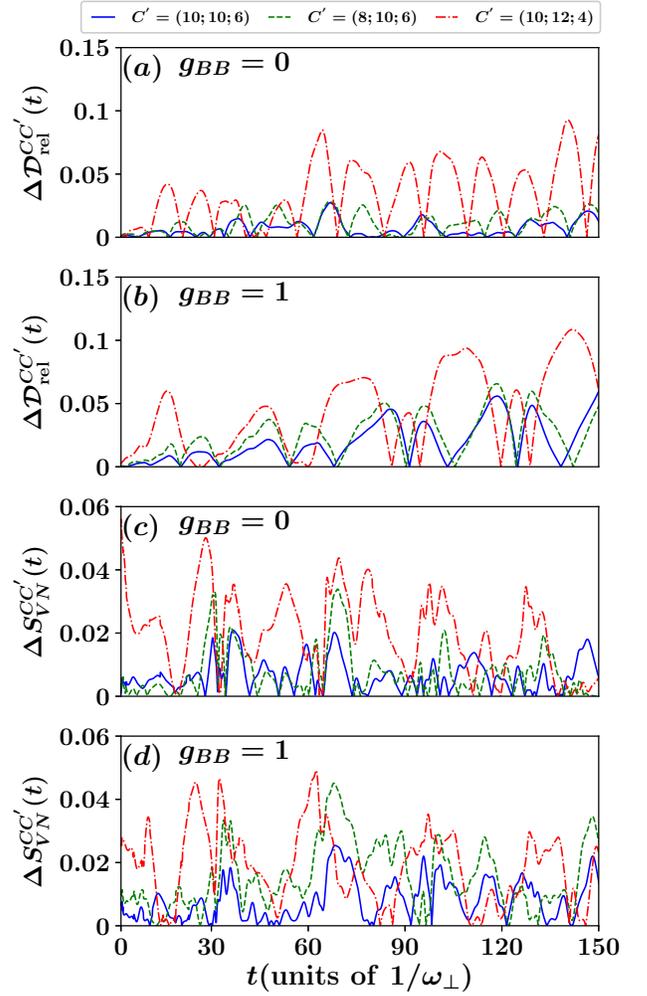}
\caption{Time-evolution of the absolute deviation of ($a$)-($b$) the relative distance $\Delta \mathcal{D}_{\rm rel}^{CC'}$ between the two bosonic impurities and ($c$)-($d$) the Von-Neumann entropy $\Delta S_{VN}^{CC'}(t)$ among the bosonic and the fermionic subsystems upon applying an interaction quench of the impurity-medium coupling from $g_{BF} = 0$ to $g_{BF} = 1$. 
The cases of [$(a)$, ($c$)] non-interacting $g_{BB} = 0$ and [($b$), ($d$)] repulsively interacting with  $g_{BB}=1$ bosonic impurities are shown. 
In all panels $C= (10;12;6)$ is kept fixed and variations of the orbital configurations denoted by $C'=(D';d'_B,;d'_F)$ are considered (see legend).}  
\label{fig:14}
\end{figure}

More specifically, the underlying Hilbert space truncation is dictated by the used orbital configuration space denoted by $C = (D;d_B;d_F)$. 
Here, $D\equiv D_B=D_F$ is the number of species functions [Eq.~({\ref{4}})] and $d_B$, $d_F$ refers to the amount of single-particle functions [Eq.~(\ref{5})] of the bosonic and the fermionic species respectively. 
Within our numerical implementation, a primitive basis consisting of a sine discrete variable representation with $600$ grid points is utilized. 
This sine discrete variable representation intrinsically introduces hard-wall boundary conditions at both edges of the numerical grid located here at $x_\pm=\pm 40$. 
Note that the aforementioned locations of the hard-walls do not affect the discussed phenomena and results since non-negligible 
portions of each subsystem density are extended up to $x_\pm=\pm 15$.  

To testify the convergence of our MB variational calculations we systematically vary the employed orbital configuration space $C=(D;d_B;d_F)$ 
until the observables of interest reach a certain level of accuracy. 
All the MB calculations presented in the main text are based on the configuration space    $C = (10;12;6)$ which has been found to provide an adequate level of numerical convergence. 
To explicitly demonstrate the degree of convergence we invoke, as case examples, the time-evolution of the relative distance between the impurities, i.e. $\mathcal{D}_{\rm rel}(t)$, and the Von-Neumann entropy $S_{VN}(t)$. 
In particular, we track their corresponding absolute deviations, namely $\Delta \mathcal{D}_{\rm rel}^{CC'}$ and $\Delta S_{VN}^{CC'}(t)$, between the $C= (10;12;6)$ and other orbital configurations designated by $C'= (D';d'_B, d'_F)$. 
Therefore, we inspect the following quantities
\begin{equation}
\begin{split}
&\Delta \mathcal{D}_{\rm rel}^{CC'}(t) = \frac{\mathcal{D}_{\rm rel}^C(t) - \mathcal{D}_{\rm rel}^{C'}(t)} {\mathcal{D}_{\rm rel}^C(t)},~~~{\rm and}~~~ 
\\&\Delta S_{VN}^{CC'}(t) = \frac{S_{VN}^{C'}(t)-S_{VN}^{C'}(t)}{S_{VN}^{C}(t)}.
\end{split}
\end{equation}

Figure~\ref{fig:14} shows the dynamics of $\Delta \mathcal{D}_{\rm rel}^{CC'}$ [Figs.~\ref{fig:14}($a$), ($b$)] and $\Delta S_{VN}^{CC'}(t)$ [Figs.~\ref{fig:14}($c$), ($d$)] for zero and finite repulsive impurity-impurity interactions following a quench of the impurity-medium coupling from $g_{BF} = 0$ to $g_{BF} = 1$. 
Apparently, an adequate degree of convergence is achieved in all cases. 
Referring to non-interacting impurities, i.e. $g_{BB}=0$, we observe that the relative difference $\Delta \mathcal{D}_{\rm rel}^{CC'}$ [$\Delta S_{VN}^{CC'}(t)$] between the $C=(10;12;6)$ and $C'=(10;10;6)$ configurations presented in Fig.~\ref{fig:14}($a$) [Fig.~\ref{fig:14}($c$)] lies below $2.5\%$ [$2\%$] throughout the evolution. 
Also, $\Delta \mathcal{D}_{\rm rel}^{CC'}$ [$\Delta S_{VN}^{CC'}(t)$] becomes at most of the order of 
$2.5 \%$ [$3.5 \%$] for $C'=(8;10;6)$ and $9\%$ [$5 \%$] when $C'=(10;12;4)$. 
Turning to the case of repulsively interacting impurities we can infer a similar convergent behavior 
with $\Delta \mathcal{D}_{\rm rel}^{CC'}$ and $\Delta S_{VN}^{CC'}(t)$ exhibiting slightly larger deviations as illustrated in Fig.~\ref{fig:14}($b$) [Fig.~\ref{fig:14}($d$)]. 
Indeed, $\Delta \mathcal{D}_{\rm rel}^{CC'}$ [$\Delta S_{VN}^{CC'}(t)$] reaches a maximum value of the order of $5.1 \%$ [$2.3 \%$] for $C'=(10;10;6)$, $6.5 \%$ [$4.2 \%$] for $C'=(8;10;6)$ and $10.3\%$ [$5 \%$] for $C'=(10;12;4)$ respectively. 
Finally, we should comment that a similar analysis has also been performed for all other post-quench impurity-medium interaction strengths and observables presented in the main text and found to be adequately converged (results not shown).

\bibliography{MB_shaking}{}
\bibliographystyle{apsrev4-1}

\end{document}